\documentclass[12pt,preprint]{aastex}

\begin{document}

\title{The Fractal Distribution of HII Regions in Disk Galaxies}

\author{N\'estor S\'anchez\altaffilmark{1} and
        Emilio J. Alfaro\altaffilmark{1}}

\altaffiltext{1}{Instituto de Astrof\'{\i}sica de Andaluc\'{\i}a,
                 CSIC, Apdo. 3004, E-18080, Granada, Spain.}

\email{nestor@iaa.es}

\begin{abstract}
It is known that the gas has a fractal structure in a wide
range of spatial scales with a fractal dimension that seems
to be a constant around $D_f \simeq 2.7$. It is expected
that stars forming from this fractal medium exhibit similar
fractal patterns. Here we address this issue by quantifying
the degree to which star-forming events are clumped. We develop,
test, and apply a precise and accurate technique to calculate
the correlation dimension $D_c$ of the distribution of HII
regions in a sample of disk galaxies. We find that the 
determination of $D_c$ is limited by the number of HII
regions, since if there are $\lesssim 100$ regions available
then a bias tending to underestimate the dimension is produced.
The reliable results are distributed in the range
$1.5 \lesssim D_c \lesssim 2.0$ with an average value $\langle
D_c \rangle = 1.81$. This corresponds to a three-dimensional
dimension of $\langle D_f \rangle = 2.73$, very similar to
the value measured in the interstellar clouds. However,
we get significant variations in the fractal dimension
among galaxies, contrary to a universal picture sometimes
claimed in literature. The fractal dimension exhibits a weak
but significant correlation with the absolute magnitude and,
to a lesser extent, with the galactic radius. The faintest
galaxies tend to distribute their HII regions in more
clustered (less uniform) patterns.
The fractal dimension for the brightest HII regions within
the same galaxy seems to be smaller than for the faintest
ones suggesting some kind of evolutionary efffect, but the
obtained correlation remains unchanged if only the brightest
regions are taken into account.
\end{abstract}



\section{Introduction}

One important issue in Astronomy is how the structure and
distribution of interstellar gas at different spatial scales
is connected to the distribution of newborn stars in the host
galaxy. It is known that the gas follows a hierarchical and
self-similar structure in a wide range of scales from $\sim
0.1$ pc to $\sim 1$ kpc \citep{Elm04,Ber07}. The fractal
dimension characterizing this self-similar structure seems
to have a nearly universal value around $D_f \simeq 2.7$
\citep{San05,San07Df}. This value is in perfect agreement with
recent simulations of compressively driven turbulence in the
Interstellar Medium done by \citet{Fed07}, who obtained $D_f
\simeq 2.6-2.7$ in their standard simulations. In principle one
would expect that newborn stars forming from the high density
peaks in these fractal regions should exhibit similar fractal
patterns \citep{Elm01}. The analysis of the young stellar
population belonging to the Gould Belt yielded a fractal
dimension $D_f = 2.68 \pm 0.04$ that is consistent with
this picture \citep{San07GB}. However, late-type stars in
the Gould Belt have a significantly larger fractal dimension
($D_f = 2.85 \pm 0.04$). There exist several possible
causes for this difference
\citep[see the discussion in][]{San07GB} but one possibility
is that these two stellar populations simply are reflecting
two different gas distributions of the parental clouds at
different spatial scales according to a multifractal scenario
\citep{Cha01,Fue06}.

To test this possibility detailed studies of fractal properties
at larger spatial scales are required. There is clear evidence
that gas also follows fractal patterns at galactic scales (i.e.
larger than the kpc). For example, the HI distribution displays
a scale-free nature in SMC \citep{Sta99}, LMC \citep{Kim03}, and
other external galaxies (both irregulars and spirals) with very
different intrinsic properties \citep{Wes99,Wil05,Beg06,Dut08}.
A fractal (or multifractal) topology can explain some galactic
properties such as flat rotation curves \citep{Pfe94} or the
Kennicutt-Schmidt star formation law \citep{Tas07}. What about
the spatial distribution of newborn stars at galactic scales?
Again, a hierarchical and self-similar picture is consistent
with the distribution of star fields and star-forming sites
on galaxy-wide scales
\citep{Fei87,Elm01,Elm03,Par03,Elm06,Fue06,Ode06,Bas07}.
However, it is not clear whether this kind of fractal
distributions are connected/related or not to some
properties of the host galaxies, such as radius, rotation,
brightness, morphology, etc. In spite of the great variety
of $D_f$ values reported in the literature for different
galaxies (for both the gas and the distribution of star
forming sites) most of the authors argue in favor of a
more or less universal picture (see the references already
mentioned). In this universal description, the constancy of
the fractal dimension is a natural consequence of the fact
that the same physical processes are structuring these systems.
However, there are some indications that the situation could
be different and the mechanism that arrange the gas at spatial
scales larger than $\sim 1$ kpc could modify the final
distribution of the star formation at such scales. The
fractal dimension of the distribution of HII regions could
be different in grand design and flocculent galaxies
\citep[as suggested by][]{Hod85}, and the brightest
galaxies could have fractal dimensions higher than
faintest ones \citep{Par03,Ode06}. On the contrary,
\citet{Fei87} do not find any correlation between the
fractal dimension and the galactic properties, but their
uncertainties are so large that the robustness of this
conclusion is questionable.

Part of the problem that prevents achieving unequivocal
conclusions lies in the great diversity of analysis
techniques used in the literature and/or the application
to not large enough samples of galaxies. This paper aims
to contribute to fill this gap by providing, firstly,
a carefully designed method that has been tested on
simulated data and that clearly establishs its accuracy
and applicability depending on the sample itself. Therefore,
this method can be used in a reliable way to investigate the
distribution of HII regions in disk galaxies. We apply this
method to the most complete sample of galaxies that we have
found in literature expecting to draw significant conclusions
regarding this matter. This paper is organized as follows.
Section~\ref{sec:teoria} explains the method used to calculate
the fractal dimension. It also analyzes the problems arising
from projection and sample size effects. Section~\ref{sec:sample}
describes the data collected from different sources on which we
apply the method in Section~\ref{sec:resultados}. The possible
correlations between the fractal dimension of the distribution
of HII regions and other galactic properties are discussed in
Section\ref{sec:correlaciones}. Finally, the main results are
summarized in Section~\ref{sec:conclusiones}.


\section{The fractal dimension of disk-like point distributions}
\label{sec:teoria}

The primary goal of this work is to calculate the fractal
dimension of the distribution of HII regions in galaxies.
One way to do this is by using the so-called correlation
dimension \citep{Gra83} which gives robust results when
dealing with distributions of points in space. Let us
consider $N$ points in space with positions ${\bf x}$.
The number of other points within a sphere of radius $r$
centered on the ${\it i}$-th point is given by
\begin{equation}
\label{eq:n}
n_i(r) = \sum_{j=1,\ j \neq i}^{N}
H \left( r- | {\bf x}_i-{\bf x}_j | \right) \ \ ,
\end{equation}
where $H(x)$ is the Heaviside step function. We can choose
$M$ different points as centers and then calculate the
commonly called correlation integral in the form
\begin{equation}
\label{eq:c}
a(r) = \frac{1}{M(N-1)} \sum_{i=1}^{M} n_i(r) \ \ .
\end{equation}
Thus, this quantity represents the probability of finding a
point within a sphere of radius $r$ centered on another point.
For a fractal set $C(r)$ scales at small $r$ as
\begin{equation}
\label{eq:dc}
C(r) \sim r^{D_c} \ \ ,
\end{equation}
being $D_c$ the correlation dimension. For a homogeneous
distribution of points in a plane we expect $D_c = 2$,
whereas if the points are distributed obeying a fractal
geometry then $D_c < 2$. When evaluating $C(r)$ for real
data the power-law behavior (eq.~\ref{eq:dc}) is valid
only within a limited range of $r$ values, even if the
distribution follows an underlying fractal law. If $r$
is of the order of the mean distance between nearest
neighbors then the distribution looks like a set of
isolated points. Furthermore, if $r$ tends to the full
data set size then boundary effects become increasingly
important. In both cases $C(r)$ deviates from the expected
power-law behavior and $D_c$ tends to be underestimated 
\citep[see][and references mentioned therein]{San07GB}.
These effects are magnified when the number of available
data decrease. We have developed an algorithm to
calculate $D_c$ in a reliable way for two-dimensional
distributions of points \citep{San07GB}. The novelty of
the algorithm lies in the implementation of objective
and suitable criteria to avoid both boundary effects and
finite-data problems at small scales. First, the algorithm
finds the boundary of the set of points by using the
procedure proposed by \citet{Edd77}, which determines the
vertices of the minimum-area convex polygon containing the
whole set of data points. Then we place circles of different
radii and evaluate $C(r)$ according to equations~(\ref{eq:n})
and (\ref{eq:c}). For this we impose the condition that all
circles must be kept inside the sample, this means that circles
are not allowed to cross the previously defined boundary. The
calculation is done for $r$ values ranging from the minimal
distance between two points to the maximum allowable value.
The number of possible samplings $M$ decreases as $r$ increases.
In each case we calculate both $C(r)$ and the corresponding
standard deviation $\sigma_C$. The correlation dimension $D_c$
is given by the slope of the best linear fit in a $\log C -
\log r$ plot. We establish a lower limit for this fit given
by the $r$ value for which $\sigma_C = C(r)$. This simple
criterion removes poorly estimated $C(r)$ values occurring
mainly at small $r$ values. An upper limit is automatically
set by the largest circle fitting into the sample. Finally,
the uncertainty associated to $D_c$ is calculated using
bootstrap techniques: we repeat the calculation on a series
of random resamplings of the data and the standard deviation
of the obtained set of $D_c$ values is taken as the error in
our estimation ($\sigma_{boot}$). This algorithm yields very
good results for distributions of points with a well-defined
fractal dimension $D_f$ (monofractals). For simulated
two-dimensional fractals with $1 < D_f < 2$ we verified that
the relation $D_c-\sigma_{boot} \lesssim D_f \lesssim D_c+
\sigma_{boot}$ is always fulfilled \citep{San07GB}. That is,
the algorithm gives unbiased fractal dimensions for sufficiently
representative samples.

What we observe on the image of an external galaxy is the
distribution of HII regions projected onto the celestial
sphere. From the inclination and position angles we can
obtain the image as projected on the mean plane of the
galaxy. One important question is the relationship between
the correlation dimension measured on this plane (let us call
it $D_c$) and the ``real" dimension of the three-dimensional
distribution of HII regions within the volume occupied by the
galactic disk ($D_f$). On the one hand, if the total thickness
of the galactic disk $Z_{disk}$ is much more smaller than the
galactic diameter $D_{gal}$, i.e. if $f \equiv Z_{disk}/D_{gal}
\ll 1$, then the system can be considered as a very thin
slice of the three-dimensional distribution. In this case,
the dimension of the three-dimensional distribution and the
two-dimensional dimension are related through the expression
\citep{Fal90}
\begin{equation}
\label{eq:slice}
D_c = D_f - 1 \ \ .
\end{equation}
On the other hand, if we have the extreme case $f \simeq 1$,
then the image should be treated as a projection of the
three-dimensional set and the expected result is \citep{Fal90}
\begin{equation}
\label{eq:pro}
D_c = \min\{2,D_f\}\ .
\end{equation}
In a previous paper \citep{San05} we showed that for real
fractals\footnote{By ``real" fractals we mean distributions of
points following an underlying fractal law but not infinite in
spatial scale or in number of points and, moreover, having some
random component.} the dimension measured on the projected image tends
to be lower than the theoretical value given by eq.~(\ref{eq:pro}).

\subsection{Projection effects}

In a disk galaxy the value of $f$ for the distribution of HII
regions would be some intermediate value between a perfect
slice ($f=0$) and a projection ($f = 1$). For a typical
galaxy like the Milky Way this value should be around $f \sim
100\ {\rm pc} / 10\ {\rm kpc} = 0.01$ \citep[e.g.][]{Pal04}.
In order to assess how the slice thickness $f$ would alter
the results we have done some numerical tests. First, we
have generated three-dimensional fractals following a
simple recipe in order to ensure a perfectly defined
fractal dimension. Within a cube of size $R$ (half of
cube height) we placed
$8$ smaller cubes (one in each octant) with size $R/L$
($L \geq 2$). Each cube is placed randomly but always within
the volume of its corresponding octant. This procedure is
repeated successively $H$ times such as at the end we get
a distribution of $8^H$ points (we have used $H=8$ which
yields $\sim 16$ millions of points). The fractal dimension
of this distribution is given by $D_f = \log 8 / \log L$.
For the extreme case $L=2$ we get a homogeneous distribution
with $D_f=3$, but if $L>2$ then $D_f < 3$. Then we took a
slice along a random plane in space and projected it onto
its mean plane. The ratio $f$ between the slice thickness
and the size of the fractal is set by hand and is kept as
a free parameter. Finally, we randomly removed points from
the fractal until reaching the desired sample size $N$ which
was also kept as a free parameter.

\begin{figure}[th]
\epsscale{0.9}
\plotone{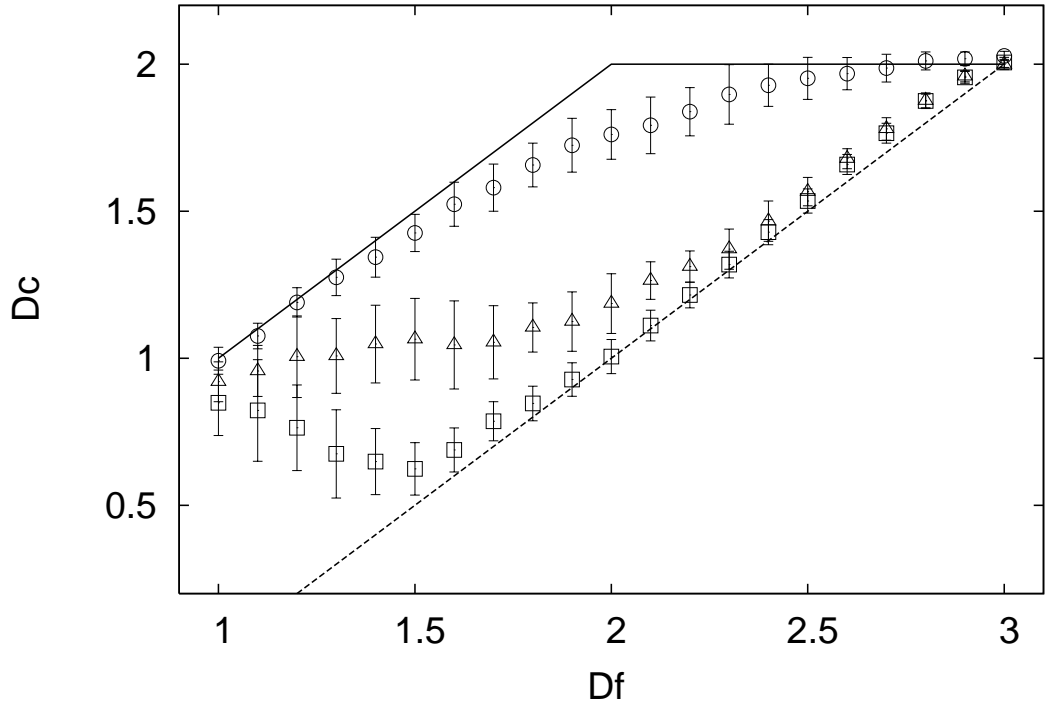}
\caption{Calculated dimension $D_c$ as a function of the
fractal dimension $D_f$ for three different values of the
slice thickness: $f=1$ (open circles), $f=10^{-2}$ (open
triangles), and $f=10^{-4}$ (open squares). The number of
points for the fractals is $N=1000$. Each point is the
result of calculating the average of $50$ random fractals,
and the bars indicate the standard deviations. The solid
and dashed lines show the theoretical results for a
projection and a thin slice, respectively.}
\label{fig:dcn1000}
\end{figure}

What we have done is to study in detail how the calculated
dimension $D_c$ depends on the original dimension $D_f$.
Figure~\ref{fig:dcn1000}
shows the results for three different values of $f$ but the
same sample size ($N=1000$ points). The error bars in this
figure come from the random component in the fractal
generation process. In general, these error bars tend
to be higher at low $D_f$ values because as $D_f$
decreases the filling factor decreases and, therefore,
the random component increases when each child cube is
placed. In other words, these error bars are related to
the random variations when generating the fractals and
not to the uncertainties in the determination of $D_c$
($\sigma_{boot}$), the latter ones always being smaller
than the former ones. The result for the case $f=1$ is
quite similar to the previously reported \citep{San05}:
the projected dimension tends to be smaller than the
theoretical value given by equation~(\ref{eq:pro}).
This is because the projection on random planes may
produce random groupings and this implies a decrease in
$D_c$. This behavior is less pronounced or even non-existent
for the cases $D_f \simeq 3$ (the projection of a homogeneous
distribution of points is again homogeneous) and $D_f \simeq 1$
(the probability of occurrence of chance groupings is low
because the filling factor is also low). In general, the
projection on the mean plane does not affect the results for
the very thin slice case ($f=10^{-4}$) with the calculated
values being very close to the theoretical values expected
for a zero-thickness slice (eq.~\ref{eq:slice}). The random
fluctuations involved in the simulations tend to ``hide" the
fractal structure, randomizing any underlying fractal pattern.
This may yield $D_c$ values slightly higher than expected for
a perfect slice. An interesting point is that when $D_f$
decreases too much, structures belonging to lower levels
of the hierarchy are so small in comparison with the slice
thickness that the resulting distribution is essentially
equivalent to a projection on the mean plane of the slice.
This change in behavior from thin slice to projection
can be clearly seen in Figure~\ref{fig:dcn1000}. The
slice-projection transition zone depends, obviously, on
both $D_f$ and $f$: as the slice thickness increases the
projection behavior appears at higher $D_f$ values. In
any case, we have obtained a nearly linear $D_c - D_f$
relationship when $D_c \gtrsim 1.5$. Moreover, this
relationship is almost independent of the exact $f$
value as long as it is around or less than the typical
value $f \simeq 0.01$ (compare, for example, cases
$f=10^{-2}$ and $f=10^{-4}$ in Figure~\ref{fig:dcn1000}).
Thus, any uncertainty in $f$ would translate into a large
uncertainty when estimating $D_f$ from $D_c$ only if $D_c
\ll 1.5$ and/or $f \gg 10^{-2}$. Otherwise the fractal
dimension would be well-determined. We will see later
that our ``reliable" results for the distribution of
HII regions in galaxies lie in the range $1.5 \leq D_c
\leq 2.0$. Table~\ref{table1} presents the resulting
values for $D_c \geq 1.5$. The results shown correspond
to the case $f=10^{-2}$ but, as mentioned before, these
results do not significantly depend on the exact $f$
value.

\subsection{Sampling size effects}

\begin{figure}[th]
\epsscale{0.9}
\plotone{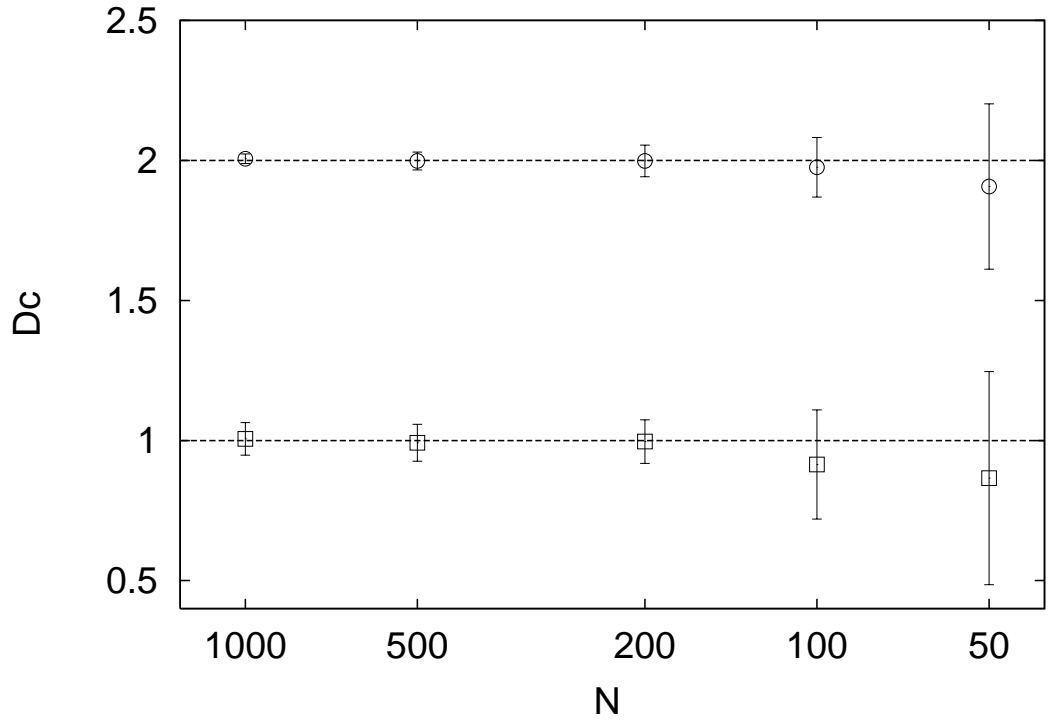}
\caption{Calculated dimension $D_c$ as a function of the
sample size $N$ for the case $f=10^{-4}$. Open circles
correspond to fractals with $D_f=3$ and open squares
to fractals with $D_f=2$. Each point is the average
of $50$ random realizations and the bars show the
standard deviations. Dashed horizontal lines show
the expected results for thin slices.}
\label{fig:dcfslice}
\end{figure}

How does the measured dimension depend on the sample size?
This is an important question because when working with
observational data many times the number of available data
points is rather small. There can be both extrinsic causes
for this (biased subsamples due to selection effects or
criteria in the detection of HII regions) and intrinsic
ones (for example, very low star formation rates would
produce a small number of clearly distinguishable HII
regions). Previously \citep{San07GB} we showed that
$D_c$ tended to be underestimated in two-dimensional
fractals having a relatively small number of points.
This effect could yield unrealistic results and/or
trends. In order to quantify this effect we repeated
the previous calculations with the same fractals and
slices but randomly removing more points until reaching
smaller sample sizes $N$. Figure~\ref{fig:dcfslice}
shows the results for $f=10^{-4}$ and two different fractal
dimensions $D_f=3$ and $D_f=2$, for which the expected values
for thin slices are $D_c=2$ and $D_c=1$, respectively
(indicated by horizontal lines). Since we are dealing
exactly with the same original data, the observed
decreasing in the average $D_c$ value with decreasing $N$
have to be attributed exclusively to sample size itself.
Thus, in general, the algorithm is able to estimate the
fractal dimension in a reliable way when applied to random
subsamples of fractal distributions of points. However, if
the sample size is too small ($N \lesssim 100$) a bias
tending to underestimate the dimension is produced.


\section{The sample of galaxies}
\label{sec:sample}

We have used VizieR\footnote{http://vizier.u-strasbg.fr}
\citep{Och00} and ADS\footnote{http://adswww.harvard.edu}
databases, in conjunction with the papers of \citet{Gar91}
and \citet{Gar02}, to search for catalogs of external
galaxies containing positions of HII regions available
in machine readable format. We found a total of 93
spiral galaxies with positions for at least 50 HII
regions. If several different catalogs were available
for the same galaxy we choose the one having the largest
amount of data. We have also included data for 8 irregular
galaxies with HII positions kindly provided by D. Hunter
\citep{Roy00}. The properties of the selected galaxies
are listed in Table~\ref{table2}. The columns contain the
following information: (1) galaxy name, (2) morphological
Hubble type, (3) and (4) position $\phi$ and inclination $i$
angles respectively (in degrees), (5) morphological type
encoded in the de Vaucouleurs scale $T$, (6) spiral arm
class $A$, (7) distance to the galaxy $D$ (in Mpc), (8)
B-band absolute magnitude $M_B$, (9) radius corresponding
to the isophotal level 25 mag/arcsec$^2$ in the B-band
$R_{25}$ (in kpc), (10) maximum rotation velocity corrected
for inclination $V_{rot}$ (in km s$^{-1}$), and (11) the
reference from which positions of HII regions were obtained
(Ref). Most of the data for spiral galaxies was taken from
HyperLeda\footnote{http://leda.univ-lyon1.fr} database
\citep{Pat03}. The arm class was taken from \citet{Elm87}.
The position and inclination angles were taken from
\citet{Gar91} and \citet{Gar02}, if available, or from
HyperLeda otherwise. All the data for irregular galaxies
come from \citet{Roy00}. Figure~\ref{fig:galaxias} shows
the deprojected distributions of HII regions for each
galaxy considered here according to the references
given in the last column of Table~\ref{table2}.


\section{The fractal dimension of the HII region distributions}
\label{sec:resultados}

\begin{figure}[th]
\epsscale{.85}
\plotone{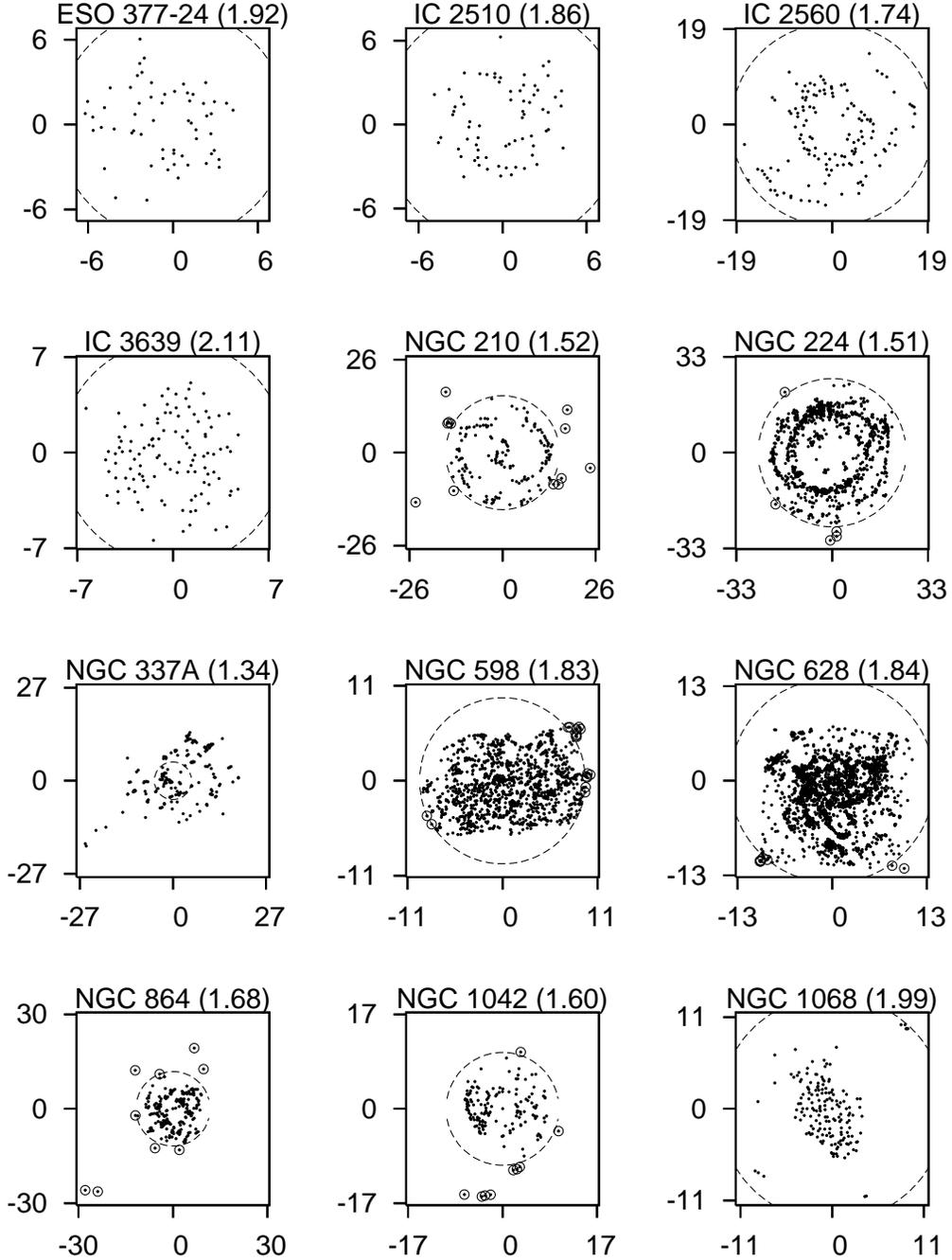}
\caption{Positions of the HII regions in the galaxy sample
according to references given in Table~\ref{table2}.
The numbers in parenthesis are the corresponding fractal
dimensions (Table~\ref{table3}).
The axis units are kpc. Dashed line circles indicate the
radius $R_{25}$. Points surrounded by circles are not
taken into account when calculating $D_c$ (see text).}
\label{fig:galaxias}
\end{figure}

It can be the case that clustering properties of HII regions
differ from region to region in the same galaxy. These
differences may be due to some bias (as, for example, if
a smaller amount of data exists in a more poorly observed
region of the galaxy) or they may be real differences (for
example, the organizing role of spiral waves will be smaller
close to the central region). Obviously, the algorithm
returns only an average estimation of $D_c$ for the
entire distribution of points. Given the sufficiently
representative sample of galaxies we expect that this
effect does not affect our main results and conclusions.
One important point that has to be taken into consideration
while estimating $D_c$ is the possible presence of outliers.
If a relatively small number of points are clearly separated
from the rest of the data then the resulting $D_c$ value
could be artificially low because of the presence of large
empty spaces. The algorithm takes this possibility into
account in a first approximation by defining the boundary.
The points that determine the boundary are not considered
in the calculation because any circle including any of
these points is automatically outside the boundary. Thus,
it does not matter if any of these points is an outlier
because it is not used in the calculation. However, we
have added an additional criterion to minimize possible
problems arising from this effect: if a small fraction
of points (we have chosen 10\% of the total sample) is
located outside $R_{25}$ then those points are removed
from the calculation. This criterion does not affect
the results if those points are not actually outliers.
The points that have been excluded from the calculation
are surrounded by circles in Figure~\ref{fig:galaxias}.

The result of calculating the correlation dimension of
the distribution of HII regions in the sample of galaxies
is shown in Table~\ref{table3}. The columns are the galaxy
name, the number of cataloged HII regions $N$, the calculated
dimension $D_c$, the uncertainty resulting from bootstrapping
$\sigma_{boot}$, and the range of spatial scales $R_{min}-R_{max}$
(in kpc) over which the HII regions are distributed.
The calculated $D_c$ values are also shown in parenthesis in
Figure~\ref{fig:galaxias}. It can be seen that $D_c$ is consistent
with the appearance of the HII region distributions. $D_c$ values
close to 2 correspond to nearly homogeneous distributions. Smaller
$D_c$ values correspond to more irregular distributions having
clumps and/or filaments separated by low density (or empty) regions.
As an example, we can compare the smallest and the largest $D_c$
values in first page of Figure~\ref{fig:galaxias}: NGC~337A
($D_c=1.34$) exhibits a much more clumpy distribution than
the relatively homogeneous galaxy NGC~1068 ($D_c=1.99$).

\begin{figure}[th]
\epsscale{.9}
\plotone{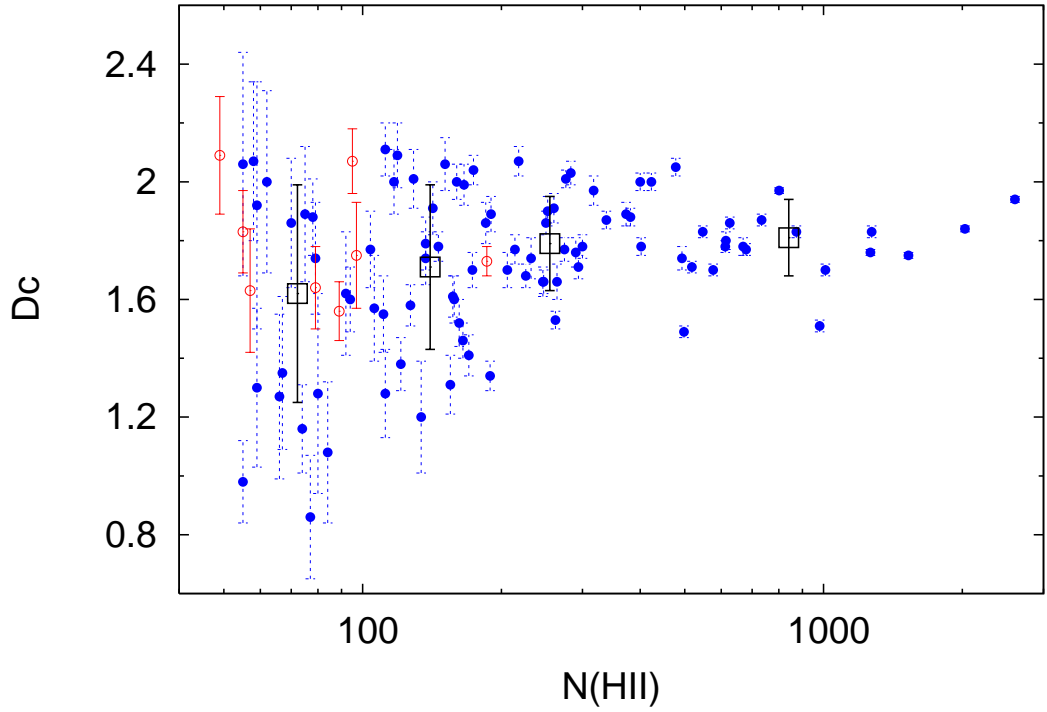}
\caption{Calculated dimension $D_c$ as a function of the
number of available data $N(\mathrm{HII})$ for spiral
galaxies (blue solid circles) and irregular galaxies
(red open circles) in the sample. The error bars indicate
the uncertainties obtained from bootstrapping. Superposed
(black squares) are the mean $D_c$ values with their
standard deviations as a function of the mean number of HII
regions for four bins having the same number of galaxies
(Table~\ref{table4}).}
\label{fig:dcnhii}
\end{figure}

Figure~\ref{fig:dcnhii}
shows the calculated correlation dimension for all the
galaxies in the sample as a function of the number of
available data points. The first thing we note is that
there exist significant differences among the galaxies:
in general, the uncertainties in the determination of
the dimension ($\sigma_{boot}$) are smaller than the
dispersion of the obtained values ($\sigma \simeq 0.3$).
This result differs considerably from the conclusion
outlined by \citet{Fei87} for their sample of 19 spiral
galaxies who argued in favor of a constant $D_c$ around
the mean value $\langle D_c \rangle = 1.68$. Two trends
are clearly visible in Figure~\ref{fig:dcnhii}. First,
the uncertainties increase as $N(\mathrm{HII})$ decreases,
consistent with the results obtained in the previous
simulations. But moreover, at small number of regions
the obtained values are substantially more spread out
toward lower values. The results were averaged in bins
of the number of HII regions with each bin having
approximately the same number of galaxies ($\sim 25$).
Table~\ref{table4} summarizes the results, where for
each bin we show mean values (also shown in
Figure~\ref{fig:dcnhii}), medians, and standard
deviations both for the fractal dimensions and for the
corresponding uncertainties. The overall trend, i.e.
the convergence of the mean value to a certain value
(in this case $\sim 1.8$) as the number of data points
increases, is very similar to the behavior described for
the simulated fractals (Figure~\ref{fig:dcfslice}). It is
clear that at least part of this trend is due to a bias in
the estimated value of $D_c$ for galaxies with small number
of HII regions. It is only when $N(\mathrm{HII}) \gtrsim 200$
that we obtain uncertainties below $\sigma_{boot} = 0.1$.
Therefore, to overcome this bias, we focus the detailed
analysis (next section) on galaxies having more than 200
HII regions (46 spiral galaxies). The average fractal
dimension in this case is $\langle D_c \rangle = 1.81$
with a standard deviations of $\sigma = 0.14$. According to
the results of Section~\ref{sec:teoria} (Table~\ref{table1}) this
corresponds to an average three-dimensional dimension of $\langle
D_f \rangle = 2.73$. Our average result is very similar 
(within the uncertainties) to the result
of \citet{Fei87} ($\langle D_c \rangle = 1.68$).
However, the average uncertainty obtained by \citet{Fei87}
is $\sim 0.3$ whereas we get an average $\sigma_{boot}$ of
$\sim 0.03$. By considering these associated uncertainties
we can say that, contrary to our results, their $D_c$ values
do not differ among themselves. Regarding the irregular
galaxies, we have only a very small number of galaxies (8)
and they have no more than 200 HII regions (the value we
are considering as the limit for an unbiased determination
of $D_c$). Thus, any conclusion based on these data should
be treated with great caution. The average fractal dimension
for irregulars is $\langle D_c \rangle = 1.79$ (similar to
that found for spirals) with a standard deviation of
$\sigma = 0.20$. The average value for $\sigma_{boot}$
is $0.14$ so that there are not significant variations
among these galaxies. The results obtained by \citet{Par03}
are spread over the range $1.2 \lesssim D_c \lesssim 2.0$,
but if we only consider their five galaxies having $\sim 200$
or more HII regions then this range narrows to $1.55 \lesssim
D_c \lesssim 1.62$. This is below our average result mostly
because our result represents a lower limit in the case of
bias due to small sample size.


\section{Dependence on galactic properties}
\label{sec:correlaciones}

At this point, we have a set of fractal dimension values
determined in a precise and accurate way and a set of
variables describing some galactic properties, such
as the position angle $\phi$, inclination $i$,
morphological type $T$, arm class $A$, distance
$D$, absolute magnitude $M_B$, radius of the isophotal
25 $R_{25}$, rotation velocity $V_{rot}$, and the number
of HII regions $N(\mathrm{HII})$. Now we proceed to examine
possible correlations between the fractal dimension and these
variables but taken into account that they represent properties
of different nature. Some of these quantities are in someway
involved in the determination of the fractal dimension and
may, therefore, introduce some bias or uncertainties into the
calculated $D_c$ values. These biases can be introduced either
by taking part directly in the calculation or through the
observation or deprojection processes. If this is the
case then spurious correlations can arise. We refer to
these as {\sl intrinsic} properties; they are $\phi$, $i$,
$D$, and $N(\mathrm{HII})$. On the other hand, the rest of
the observed variables are related to galactic properties
but they do not participate directly in the determination
of $D_c$; these {\sl extrinsic} variables are $T$, $A$,
$M_B$, $R_{25}$, and $V_{rot}$. We also analyze a new
variable that we have called the average surface density
of star forming regions, given by the number of star
forming regions divided by the square of the radius
($N(\mathrm{HII})/R_{25}^2$).

\subsection{Analysis method}

The main goal is to identify possible relationships between
the degree of clumpiness of HII regions in galaxies and their
{\sl intrinsic} and/or {\sl extrinsic} properties, in the above
sense. So we have to suggest a model of dependence and evaluate
its ``goodness" via some statistical test defined for such
purpose. We propose, in a first approximation, a linear model
linking the variables and we choose as statistical criterion
the $AIC$ \citep[Akaike's Information Criterion,][]{Aka74},
defined as $AIC=2k-2\ln(L)$ where $L$ is the maximized likelihood
value and $k$ is the number of estimable parameters in the model.
Let $RSS$ be the residual sum of squares and $n$ the number of
observations, then the $AIC$ becomes $AIC=2k+n\left[\ln(2\pi
RSS/n)+1\right]$. Increasing the number of free parameters to
be estimated improves the goodness of fit, regardless of the
number of free parameters in the data generating process. Hence
$AIC$ not only rewards goodness of fit, but also includes a
penalty that is an increasing function of the number of
estimated parameters. This penalty discourages overfitting.
The preferred model is the one with the lowest $AIC$ value.
Once the best model has been selected according to this
criterion we perform the $t$ and $F$ tests to infer the
significance of the coefficients of the model and of the
overall fit. This analysis is performed in the {\bf R}
environment for statistical computing \citep{R08}.

\subsection{Intrinsic variables}

We begin with the intrinsic variables ($\phi$, $i$, $D$, and
$N(\mathrm{HII})$) and considering first the whole sample of
galaxies. Starting from a linear regression model based on $n$
variables we apply a stepwise regression process to select the
subset of variables providing the best model. The analysis
yields the position angle of the major axis and the number
of HII regions as the most significant variables. Thus, the
best model to explain $D_c$ variations from these variables
has the form $D_c = a\, \phi + b\, N(\mathrm{HII}) + c$.
However, both the $t$-test for the regression coefficients
and the $F$-test for the overall fit indicate no significance
at 95\% confidence level. This result supports our previous
finding in simulated fractal distributions: the number of
HII regions in a galaxy affects the quality of the estimated
fractal dimension by introducing a systematic bias and a noise.
The presence of the variable $N(\mathrm{HII})$ in the model is
simply the result of the existence of this bias.
If we exclude those objects with $N(\mathrm{HII}) < 200$ and
repeat the analysis, we obtain that any variable is able to
significantly explain the observed variance in the fractal
dimension.

\subsection{Extrinsic variables}

\begin{figure}[th]
\plotone{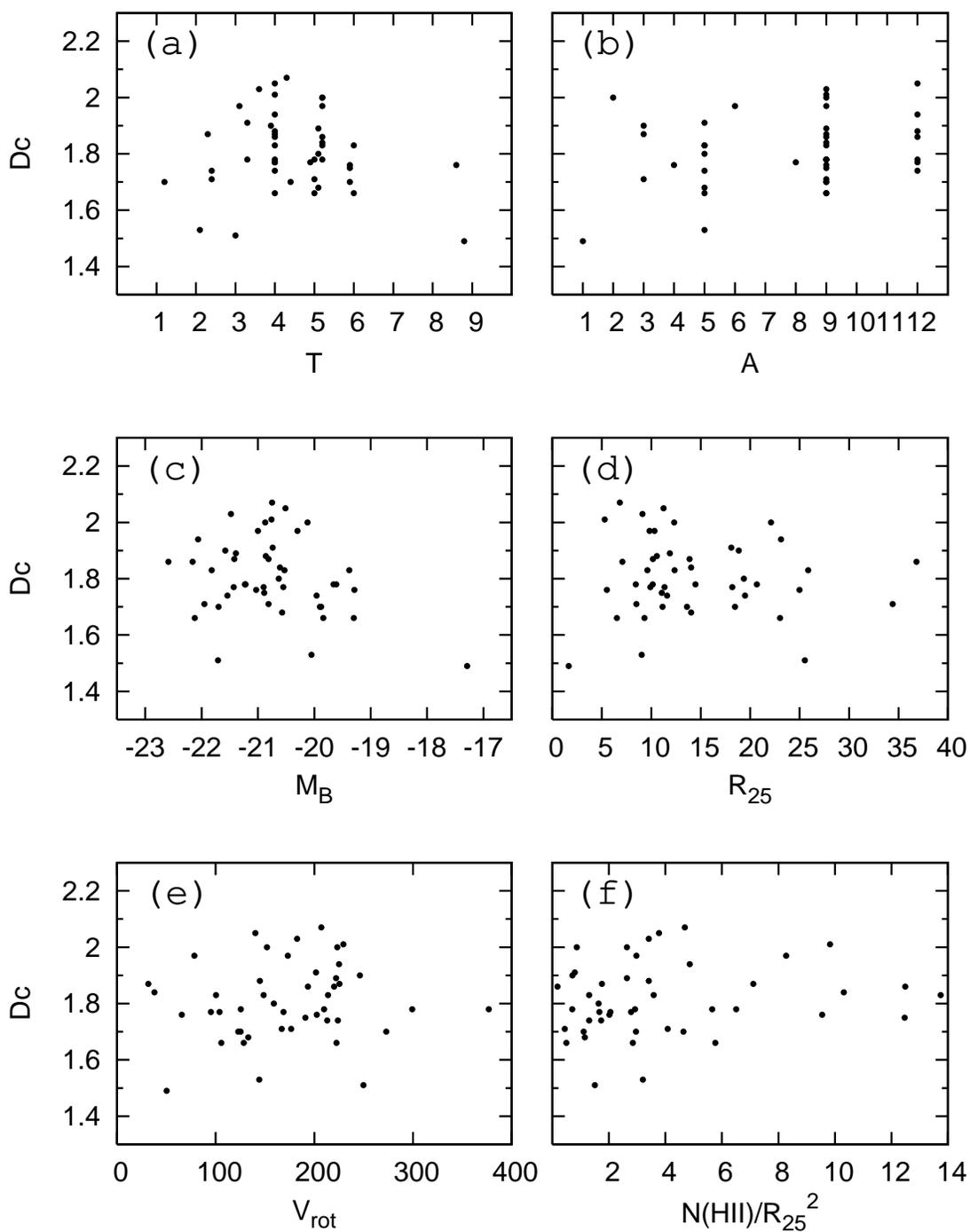}
\caption{Calculated dimension $D_c$ as a function of (a)
morphological type $T$, (b) arm class $A$, (c) absolute magnitude
$M_B$, (d) radius $R_{25}$, (e) rotation velocity $V_{rot}$,
and (f) average surface density of star forming regions
$N(\mathrm{HII})/R_{25}^2$. Only galaxies having
$N(\mathrm{HII}) > 200$ have been plotted.}
\label{fig:extrinsecas}
\end{figure}

Now we will consider the variables $T$, $A$, $M_B$, $R_{25}$,
$V_{rot}$, and $N(\mathrm{HII})/R_{25}^2$, and we will restrict
the analysis to galaxies having $N(\mathrm{HII}) > 200$.
Figure~\ref{fig:extrinsecas}
shows $D_c$ as a function of each one of these variables. First
we construct an initial linear model incorporating all of these
variables. The application of the task {\sl step} (within the
package {\sl stats} of {\bf R}) yields that the variables $M_{B}$
and $R_{25}$ represent the best liner model fitting the data,
this is
\begin{equation}
\label{eq:fit}
D_c= -(0.069 \pm 0.025) M_{B} -(0.006 \pm 0.003) R_{25} +0.456 \ \ .
\end{equation}
The overall fit gives $F=4.0$ that for degrees of freedom 2 and 43
is significant at the confidence level of 95\%. The regression
coefficients of the variables $M_{B}$ and $R_{25}$ are significant
at the confidence levels of 99\% and 95\% (marginally significant),
respectively.

According to our results $D_c$ correlates with $M_{B}$ and, to a
lesser extent, with $R_{25}$. Brightest galaxies have fractal
dimensions higher than faintest ones (Figure~\ref{fig:extrinsecas}c).
A slight trend is apparent in Figure~\ref{fig:extrinsecas}c in
which the correlation with $M_{B}$ appears to saturate for galaxies
brighter than $M_{B} \simeq -21$, but the large scatter in this
$M_{B}$ interval prevents us from drawing a general conclusion.
\citet{Fei87} find a pure scatter diagram when comparing $D_c$ and
$M_B$ which, together with the lack of correlation with other
properties is used as argument favoring a universal $D_c$ value
in spiral galaxies. The dependence on $M_B$ that we observe is
probably hidden by their imprecise determination of $D_c$
(uncertainties of the order of $\sim 0.3$). In contrast,
\citet{Par03} report the same trend in their sample of
dwarf irregulars, i.e. the highest fractal dimensions for
the brightest galaxies. They applied a correction factor
to the calculated $D_c$ to take account of differences in
the number of HII regions. However, their results have to be
taken with caution because, as mentioned before, it is a
little disturbing that they have only five galaxies with
$N(\mathrm{HII}) \gtrsim 200$ for which the correlation
does not exist at all. \citet{Ode06} analyzed in detail the
distribution of resolved young stars in star-forming dwarf
galaxies. Unfortunately she used only four galaxies, but
she obtained that the brightest galaxy had the highest
fractal dimension and the faintest galaxy the lowest.
The interpretation given by \citet{Par03} is that
a small $D_c$ value implies a higher intragalactic
gas porosity which reduces the (massive) star formation
rate \citep{Sil97}. The same argument remains valid for
spiral galaxies if the presence of density waves does
not greatly affect the star formation process \citep{Elm86}.
For the irregular galaxies in our sample, we find that the
average fractal dimension ($\simeq 1.79$) is similar within
the uncertainties to that found for spirals with
$N(\mathrm{HII}) \gtrsim 200$ ($\simeq 1.81$).
Moreover, it is higher than that for spirals with
$N(\mathrm{HII}) < 200$ ($\simeq 1.65$) although,
statistically, there is no significative difference.
Curiosity prompted us to check what happens when irregular
galaxies are included in the previous statistical analyses,
and what we find is that the correlation disappears.
Dwarf irregular galaxies are characterized by fractal
dimensions similar to those of the brightest spiral
galaxies, so that any possible correlation between
$D_c$ and $M_B$ is removed when both types of
galaxies are considered. Even if we try to correct
the small sample size bias in the irregulars, the
resulting $D_c$ values would be higher and this
effect would be enhanced. Therefore, in spite of
the small number of irregular galaxies studied,
we can conclude that if $D_c$ is a function of
$M_B$ in irregulars \citep[as suggested by][]{Par03}
this correlation does not follow the same law as for
spirals; and the fractal dimension for irregulars
is, on average, higher than for spirals with the same
absolute magnitude.

\subsection{Possible evolutionary effects}

\begin{figure}[th]
\plotone{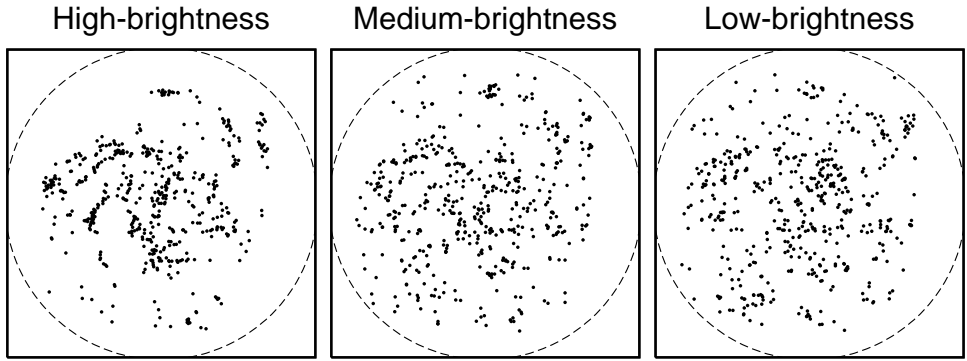}
\caption{Positions of the HII regions in the galaxy
NGC~6946 for high, medium, and low brightness regions
(see text). The resulting fractal dimensions were
$D_c = 1.64$ (high-brightness), $D_c = 1.82$
(medium-brightness), and $D_c = 1.79$ (low-brightness).}
\label{fig:NGC6946}
\end{figure}

If we assume that star formation takes place mainly along
the spiral arms then the inter-arm regions should typically
exhibit older HII regions. This kind of evolutionary effect
have been proposed to explain variations in the luminosity
functions of HII regions in the spiral arms versus the
inter-arm regions \citep[e.g.][]{Oey98}.
We wonder whether this kind of effect could introduce some
bias because in principle one would expect that only
the brightest regions are visible for the farthest
(and faintest) galaxies. However, we didn't detect any
correlation between the calculated $D_c$ value and the
galaxy's distance so that our results would seem to be
unaffected by this possible effect.
In any case, we have analyzed the dependence of $D_c$ on
the brightness of the HII regions for those galaxies
with enough available data. We first selected galaxies
from the sample for which we have data on HII region
brightness and a sample size large enough
($N(\mathrm{HII}) > 600$) to calculate $D_c$ in a
reliable way for the brightest 1/3 of the HII regions.
There are a total of 9 galaxies fulfilling these
requiriments:
NGC~628, NGC~3344, NGC~3486, NGC~3631, NGC~3726,
NGC~4254, NGC~4303, NGC~4321, and NGC~6946
\citep[the reference for this data is][]{Bra06}.
We divided the sample in three equal subsamples
ordered in descending brightness: the brightest
HII regions (high-brightness), the medium bright
ones (medium-brightness), and the faintest ones
(low-brightness).
Figure~\ref{fig:NGC6946} shows, as an illustrative
example, the resulting distributions in NGC~6946
for which we have little more than $\sim 500$ HII
regions in each case.  
We recalculated $D_c$ for each case and the results
are shown in Table~\ref{table5}.
The general trend is that $D_c$ is smaller for
the brightest HII regions than for the rest of
the data. This can be seen by eye from inspection
of Figure~\ref{fig:NGC6946} where the left-side
panel exhibits, clearly, a more clumpy morphology.
The only exceptions to this general trend are
NGC~4254 that exhibits the opposite behavior and
NGC~4303 that does not show any significant
variation. The rest of the galaxies have a
smaller $D_c$ value for the brightest HII
regions. These regions should reflect, in a
first approximation, the initial distribution
of star-forming regions in each galaxy. It
seems likely from our results that some kind
of evolutionary effect tends to randomize
(homogenize) in some degree the initial
distributions of HII regions.

Does this result affect the correlations we found
in the previous sections? We have repeated the
analysis replacing the full sample $D_c$ values
by the high-brightness values for the 9 galaxies
in Table~\ref{table5}. First we use all the galaxies
having more than $200$ HII regions (46 galaxies).
In this case, the analysis yields exactly the same
variables for the best fit (eq.~\ref{eq:fit}) with
nearly the same $F$ value ($4.1$) and a confidence level
of 99\% for the overall fit. The coefficients for $M_{B}$
and $R_{25}$ are $-0.069$ and $-0.007$ with confidence
levels of 99\% and 95\%, respectively. We have also
tested what happens if we take only the galaxies
having more than $200$ and less than $600$ HII
regions and the high-brightness values for the
9 galaxies in Table~\ref{table5}. The idea behind
this selection criterion is to avoid galaxies having
a ``mixed" (bright and faint) population of HII regions.
Again, the analysis yields $M_{B}$ and $R_{25}$ as the
variables providing the best fit to the data. The fit
gives $F=4.6$ that for degrees of freedom 2 and 37 is
significant at the 95\%. The coefficients obtained in
this case for $M_{B}$ and $R_{25}$ are $-0.079$ and
$-0.006$ with confidence levels of 99\% and 90\%,
respectively.
Summarizing, when only the bright HII regions are
considered the trend shown in eq.~(\ref{eq:fit})
is preserved, i.e. the correlation strength with
$M_{B}$ remains the same (or even increases) whereas
the correlation with $R_{25}$ is, in the best scenario,
marginally significant.


\section{Conclusions}
\label{sec:conclusiones}

The study of the distribution of HII regions in galaxias may
provide important clues to understanding the processes involved
in the star formation at galactic scales. Our approach in this
work has been apply a precise and accurate technique to calculate
the degree of self-similar clumpiness (the correlation dimension
$D_c$) in a large enough sample of disk galaxies. All the obtained
dimensions are shown in Table~\ref{table3}. Our most reliable results
are distributed in the range $1.5 \lesssim D_c \lesssim 2.0$ with an
average value $\langle D_c \rangle = 1.81$. This value corresponds
to a three-dimensional dimension of $\langle D_f \rangle = 2.73$.
The similarity between this value and the value $D_f \simeq 2.7 \pm
0.1$ measured in interstellar molecular clouds \citep{San07Df} may
seem, in principle, unexpected given the very different spatial
scales involved. There are many energy sources that can drive
the turbulence and structure the interstellar medium \citep{Elm04},
but this structure can be modified at a galactic level due to the
action of other physical processes acting at this larger spatial
scales. The fractal behavior we have observed span a range of
scales from $\sim 10$ pc to $\sim 25$ kpc, and we have not found
any dependence on the spatial scales involved.

The uncertainties of the most reliable $D_c$ values are small
enough ($\sim 0.03$ on average) to allow us to conclude that
there are significant variations among galaxies. This conclusion
is incompatible with a universal picture in which the fractal
dimension is approximately a constant for the spiral galaxies
\citep{Fei87}. We have found that $D_c$ exhibits a weak but
significant correlation with $M_{B}$ and, to a lesser extent,
with $R_{25}$. The faintest galaxies tend to distribute the
HII regions in a more clumpy or clustered way. It is the
first time that this behavior is reported in spiral galaxies,
but it is in agreement with similar results suggested for
irregular galaxies \citep{Par03,Ode06}.
Moreover, the fractal dimension for the brightest HII regions
tends to be smaller than for the faintest ones. This behavior
could be the result of some kind of evolutionary effect that
tends to randomize in some degree the initial distribution
of HII regions \citep{Oey98}. However, if only the brightest
HII regions are taken into account the observed correlation
with $M_{B}$ and $R_{25}$ remains unchanged.
We have not found correlations with other galactic properties
(morphological type, spiral arm class, rotation velocity, and
surface density of HII regions).
There are many galactic properties from which $D_c$ could depend,
such as star formation activity, mass, age, metallicity, or a
combination of them. It has been suggested that the fractal
dimension of the distribution of star-forming sites could be
increased during the star formation process
\citep[for example,][and this work]{Fue06}.
A more complete analysis including more galactic variables
and a wider and more diverse sample of galaxies would be
necessary in order to obtain a clearer picture.

\acknowledgments
We acknowledge D.~Hunter for providing the data on irregular
galaxies. We also thank an anonymous referee for his/her
comments, which improved this paper and led us to add
Section~5.4 to an earlier version of this paper.
This research has made use of NASA's Astrophysics
Data System and of HyperLeda and VizieR databases.
We acknowledge financial support from MEC of Spain through grant
AYA2007-64052 and from Consejer\'{\i}a de Educaci\'on y Ciencia
(Junta de Andaluc\'{\i}a) through TIC-101.



\clearpage

\begin{deluxetable}{ccc}
\tablecolumns{3}
\tablewidth{0pt}
\tablecaption{Results for the simulated fractals
for the $f=10^{-2}$ case (see text)\label{table1}}
\tablehead{
\colhead{$D_f$} & \colhead{$D_c$} &
\colhead{$\sigma_{boot,ave}$\tablenotemark{1}}
}
\startdata
2.40 & 1.47 & 0.02 \\
2.45 & 1.52 & 0.02 \\
2.50 & 1.57 & 0.02 \\
2.55 & 1.63 & 0.02 \\
2.60 & 1.68 & 0.02 \\
2.65 & 1.73 & 0.02 \\
2.70 & 1.78 & 0.02 \\
2.75 & 1.83 & 0.02 \\
2.80 & 1.88 & 0.02 \\
2.85 & 1.92 & 0.02 \\
2.90 & 1.96 & 0.02 \\
2.95 & 1.98 & 0.02 \\
3.00 & 2.00 & 0.02
\enddata
\tablenotetext{1}{Average of the calculated uncertainties,
as determined from bootstrapping}
\end{deluxetable}

\clearpage

\begin{deluxetable}{llrrrcrrrrc}
\tablecolumns{10}
\tablewidth{0pt}
\tablecaption{Properties of the galaxies in the sample\label{table2}}
\tablehead{
\colhead{Name} &
\colhead{type} &
\colhead{$\phi$} &
\colhead{$i$} & 
\colhead{$T$} &
\colhead{$A$} &
\colhead{$D$} &
\colhead{$M_B$} &
\colhead{$R_{25}$} &
\colhead{$V_{rot}$} & 
\colhead{Ref.}
}
\startdata
\cutinhead{Spirals}
ESO 377-24 & Sc   & 100 & 35 & 5.0 &\nodata&  39.5 & -20.10 &  8.14 & 129.9 & (1) \\
IC 2510    & SBb  & 152 & 60 & 2.5 &\nodata&  37.5 &\nodata &  8.68 & 133.7 & (1) \\
IC 2560    & SBb  &  42 & 57 & 3.3 &\nodata&  39.2 & -21.16 & 20.15 & 195.9 & (1) \\
IC 3639    & SBbc & 105 & 20 & 4.1 &\nodata&  44.9 & -20.48 &  8.21 & 215.1 & (1) \\
NGC 210    & SABb & 162 & 57 & 3.1 &     6 &  22.0 & -20.41 & 15.89 & 154.3 & (2) \\
NGC 224    & Sb   &  38 & 77 & 3.0 &\nodata&   1.0 & -21.71 & 25.53 & 249.8 & (3) \\
NGC 337A   & SABd &   8 & 56 & 7.9 &     2 &  14.2 & -18.65 &  5.45 &  39.3 & (2) \\
NGC 598    & Sc   &  23 & 55 & 6.0 &     5 &   1.1 & -19.38 &  9.62 & 100.4 & (4) \\
NGC 628    & Sc   &  23 &  8 & 5.2 &     9 &   9.8 & -20.61 & 14.02 &  38.0 & (2) \\
NGC 864    & SABc &  24 & 48 & 5.1 &     5 &   9.8 & -20.57 & 14.02 & 133.0 & (2) \\
NGC 1042   & SABc &   6 & 16 & 6.0 &     9 &  21.8 & -20.23 & 11.82 & 144.7 & (2) \\
NGC 1068   & Sb   &  84 & 29 & 3.0 &     3 &  18.0 & -21.38 & 10.16 & 320.9 & (2) \\
NGC 1073   & SBc  & 163 & 30 & 5.3 &     5 &  15.3 & -19.91 & 13.55 &  38.6 & (2) \\
NGC 1097   & SBb  & 130 & 50 & 3.3 &    12 &  16.4 & -21.22 &  8.43 & 299.3 & (5) \\
NGC 1300   & Sbc  & 101 & 49 & 4.0 &    12 &  15.2 & -20.93 & 23.51 & 167.1 & (2) \\
NGC 1566   & SABb &  41 & 32 & 4.0 &    12 &  20.2 & -21.43 & 18.18 & 104.0 & (1) \\
NGC 1672   & Sb   & 170 & 34 & 3.3 &     5 &  17.4 & -20.74 & 18.08 & 201.7 & (5) \\
NGC 1808   & Sa   & 128 & 48 & 1.2 &\nodata&  15.0 & -19.90 & 13.62 & 122.9 & (1) \\
NGC 2775   & Sab  & 160 & 41 & 1.7 &     3 &  10.9 & -20.60 &  8.62 & 291.4 & (2) \\
NGC 2805   & SABc & 116 & 39 & 6.9 &     5 &  19.2 & -20.84 & 11.88 &  70.7 & (2) \\
NGC 2985   & Sab  & 180 & 36 & 2.3 &     3 &  28.1 & -20.81 & 13.87 & 225.4 & (2) \\
NGC 2997   & SABc & 100 & 40 & 5.1 &     9 &  22.6 & -21.39 & 11.89 & 222.1 & (6) \\
NGC 3031   & Sab  & 150 & 58 & 2.4 &    12 &  13.1 & -21.54 & 19.48 & 223.8 & (7) \\
NGC 3081   & S0-a &  91 & 39 & 0.0 &     6 &   5.3 & -19.99 & 16.53 &  99.9 & (1) \\
NGC 3169   & Sa   &  45 & 57 & 1.1 &     2 &  31.8 & -20.34 & 12.21 & 321.0 & (8) \\
NGC 3184   & SABc &  90 & 13 & 5.9 &     9 &  17.7 & -19.88 & 11.14 & 125.3 & (2) \\
NGC 3227   & SABa & 157 & 68 & 1.4 &     7 &  10.9 & -20.12 & 11.77 & 130.0 & (2) \\
NGC 3344   & Sbc  & 164 & 28 & 4.0 &     9 &  17.5 & -19.66 & 10.14 & 209.9 & (2) \\
NGC 3351   & Sb   &  13 & 39 & 3.0 &     6 &   9.9 & -20.23 &  9.57 & 176.7 & (2) \\
NGC 3359   & Sc   & 171 & 52 & 5.2 &     5 &  11.8 & -20.53 & 12.35 & 148.6 & (9) \\
NGC 3367   & Sc   & 100 & 40 & 5.2 &     9 &  18.0 & -21.39 & 10.68 & 172.8 & (8) \\
NGC 3368   & SABa &   5 & 55 & 1.8 &     8 &  44.3 & -20.90 & 18.02 & 194.0 & (2) \\
NGC 3393   & SBa  &  80 & 27 & 1.1 &\nodata&  13.4 & -20.83 & 16.48 & 168.8 & (1) \\
NGC 3486   & Sc   &  80 & 42 & 5.2 &     9 &  51.7 & -19.61 & 14.46 & 125.4 & (2) \\
NGC 3631   & Sc   & 150 & 22 & 5.2 &     9 &  11.6 & -21.00 &  9.84 &  78.4 & (2) \\
NGC 3660   & Sbc  & 120 & 37 & 3.9 &     2 &  19.9 & -21.16 & 10.74 & 230.7 & (8) \\
NGC 3726   & Sc   &  14 & 49 & 5.1 &     5 &  52.0 & -20.63 & 19.36 & 159.0 & (2) \\
NGC 3783   & SBa  & 137 & 30 & 1.3 &     9 &  15.4 & -21.09 & 11.78 & 126.7 & (1) \\
NGC 3810   & Sc   &  32 & 48 & 5.2 &     2 &  38.9 & -20.12 & 12.31 & 152.1 & (2) \\
NGC 3982   & SABb &   7 & 21 & 3.2 &     2 &  15.2 & -19.83 &  7.35 & 190.4 & (8) \\
NGC 4030   & Sbc  &  19 & 40 & 4.0 &     9 &  19.8 & -20.76 &  5.30 & 229.4 & (2) \\
NGC 4051   & SABb & 132 & 40 & 4.0 &     5 &  21.1 & -19.96 & 11.60 & 213.1 & (2) \\
NGC 4123   & Sc   & 141 & 42 & 5.0 &     9 &  13.1 & -19.84 &  9.31 & 128.5 & (2) \\
NGC 4151   & SABa &  50 & 21 & 2.1 &     5 &  19.5 & -20.05 &  9.04 & 144.3 & (2) \\
NGC 4254   & Sc   &  61 & 30 & 5.2 &     9 &  16.9 & -22.59 &  7.08 & 193.5 & (2) \\
NGC 4258   & SABb & 157 & 72 & 4.0 &\nodata&  35.8 & -21.04 & 26.21 & 208.0 & (10)\\
NGC 4303   & Sbc  & 135 & 29 & 4.0 &     9 &   9.7 & -21.82 & 25.88 & 213.8 & (2) \\
NGC 4321   & SABb & 153 & 32 & 4.0 &    12 &  23.1 & -22.06 & 23.13 & 225.0 & (2) \\
NGC 4395   & Sm   & 127 & 18 & 8.8 &     1 &   2.7 & -17.29 &  1.66 &  50.3 & (2) \\
NGC 4487   & Sc   &  74 & 46 & 5.9 &     5 &  14.7 & -19.56 &  7.38 & 119.7 & (2) \\
NGC 4507   & Sab  &  60 & 33 & 1.9 &     5 &  47.7 & -20.92 &  9.57 &\nodata& (1) \\
NGC 4535   & Sc   & 181 & 44 & 5.0 &     9 &  29.0 & -21.95 & 34.41 & 176.6 & (2) \\
NGC 4548   & Sb   & 150 & 37 & 3.1 &     5 &   8.5 & -20.81 &  6.83 & 163.2 & (2) \\
NGC 4579   & SABb &  89 & 39 & 2.8 &     9 &  22.9 & -21.68 & 16.71 & 263.1 & (2) \\
NGC 4593   & Sb   &  80 & 44 & 3.0 &     5 &  35.3 & -20.81 & 12.33 & 295.8 & (1) \\
NGC 4602   & SABb &  96 & 65 & 4.3 &\nodata&  36.4 & -20.75 &  6.82 & 207.0 & (1) \\
NGC 4618   & SBm  &  40 & 58 & 8.6 &     4 &  10.7 & -19.29 &  5.51 &  65.6 & (2) \\
NGC 4639   & Sbc  & 127 & 55 & 3.5 &     2 &  15.4 & -19.15 &  6.44 & 163.5 & (5) \\
NGC 4689   & Sc   & 163 & 31 & 4.7 &     3 &  24.6 & -20.69 & 13.66 & 135.1 & (2) \\
NGC 4699   & SABb &  47 & 47 & 3.0 &     3 &  19.9 & -21.49 & 13.76 & 256.9 & (8) \\
NGC 4725   & SABa &  32 & 54 & 2.1 &     6 &  19.5 & -21.77 & 27.50 & 225.4 & (2) \\
NGC 4736   & Sab  & 118 & 37 & 2.4 &     3 &   7.5 & -20.81 &  8.49 & 166.9 & (2) \\
NGC 4939   & Sbc  &  19 & 56 & 4.0 &    12 &  44.3 & -22.16 & 36.82 & 220.1 & (1) \\
NGC 4995   & SABb & 100 & 50 & 3.1 &     6 &  25.0 & -20.58 &  8.81 & 163.2 & (8) \\
NGC 5033   & Sc   & 171 & 62 & 5.2 &     9 &  15.4 & -20.87 & 22.10 & 223.3 & (5) \\
NGC 5194   & Sbc  & 170 & 20 & 4.0 &    12 &  10.0 & -20.51 & 11.25 & 140.2 & (11)\\
NGC 5247   & SABb & 170 & 38 & 4.1 &     9 &  18.9 & -21.16 & 14.59 &  95.6 & (2) \\
NGC 5248   & SABb & 146 & 61 & 4.0 &    12 &  17.8 & -20.86 & 10.57 & 144.8 & (2) \\
NGC 5334   & Sc   &  18 & 45 & 5.1 &     2 &  20.5 & -19.13 & 10.07 & 133.7 & (2) \\
NGC 5364   & Sbc  &  37 & 52 & 4.0 &     9 &  18.9 & -20.88 & 10.44 & 170.1 & (6) \\
NGC 5371   & Sbc  &   7 & 39 & 4.0 &     9 &  39.8 & -22.12 & 23.03 & 222.5 & (2) \\
NGC 5427   & SABc &  90 & 30 & 5.0 &     9 &  39.4 & -21.23 & 20.68 & 376.9 & (1) \\
NGC 5457   & SABc &  39 & 18 & 5.9 &     9 &   7.2 & -21.03 & 25.00 & 202.4 & (12)\\
NGC 5474   & Sc   &  85 & 50 & 6.0 &     2 &   7.4 & -16.06 &  2.58 &  22.6 & (2) \\
NGC 5643   & Sc   &  70 & 30 & 4.9 &\nodata&  14.8 & -20.90 & 11.34 & 168.7 & (1) \\
NGC 5681   & Sb   & 152 & 56 & 3.5 &\nodata& 114.9 & -21.39 & 14.49 & 187.9 & (6) \\
NGC 5850   & Sb   & 108 & 47 & 3.1 &     8 &  37.7 & -21.48 & 18.35 & 117.4 & (2) \\
NGC 5921   & Sbc  & 149 & 36 & 4.0 &     8 &  22.6 & -20.55 &  9.92 &  95.2 & (2) \\
NGC 5964   & SBcd & 147 & 42 & 6.9 &     2 &  22.2 & -18.89 & 11.05 & 120.8 & (2) \\
NGC 6070   & Sc   &  65 & 60 & 6.0 &     9 &  30.0 & -21.09 & 15.21 & 201.4 & (6) \\
NGC 6118   & Sc   &  57 & 70 & 5.9 &\nodata&  23.5 & -20.80 & 15.20 & 163.1 & (6) \\
NGC 6140   & Sc   &  76 & 32 & 5.6 &     2 &  17.1 & -19.15 &  5.21 & 169.3 & (2) \\
NGC 6221   & Sc   &  25 & 45 & 4.9 &\nodata&  18.4 & -21.63 & 12.90 & 162.9 & (1) \\
NGC 6300   & SBb  & 108 & 52 & 3.1 &     6 &  12.8 & -20.30 & 10.32 & 173.2 & (5) \\
NGC 6384   & SABb &  40 & 47 & 3.6 &     9 &  25.5 & -21.48 &  9.11 & 182.5 & (6) \\
NGC 6814   & SABb & 176 & 22 & 4.0 &     9 &  22.6 & -21.42 & 10.16 &  31.8 & (13)\\
NGC 6946   & SABc &  64 & 34 & 5.9 &     9 &   6.7 & -20.89 & 11.07 & 190.9 & (2) \\
NGC 7314   & SABb & 178 & 64 & 4.0 &     2 &  18.8 & -20.48 & 11.49 & 150.5 & (5) \\
NGC 7331   & Sbc  & 170 & 77 & 3.9 &     3 &  14.1 & -21.58 & 18.85 & 246.1 & (14)\\
NGC 7479   & SBbc &  22 & 51 & 4.4 &     9 &  34.9 & -21.70 & 18.47 & 273.0 & (15)\\
NGC 7552   & Sab  & 181 & 37 & 2.4 &\nodata&  20.2 & -20.45 & 11.38 & 208.1 & (6) \\
NGC 7590   & Sbc  &  25 & 65 & 4.1 &\nodata&  19.8 & -20.08 &  8.78 & 182.5 & (1) \\
NGC 7741   & SBc  & 162 & 43 & 6.0 &     5 &  12.4 & -19.30 &  6.53 & 105.7 & (2) \\
\cutinhead{Irregulars}
DDO 47   & IB & 100.0 & 66.0 & 9.8 &\nodata& 2.8 & -14.15 & 1.25 & 62 & (16)\\
DDO 50   & I  &  25.0 & 50.2 & 9.8 &\nodata& 3.2 & -16.93 & 3.70 & 41 & (16)\\
DDO 154  & I  &  49.5 & 59.5 & 9.9 &\nodata& 4.0 & -14.52 & 1.75 & 48 & (16)\\
DDO 168  & I  & 149.0 & 66.0 & 9.9 &\nodata& 3.5 & -15.43 & 1.85 & 55 & (16)\\
IC 10    & IB & 137.5 & 54.5 & 9.9 &\nodata& 1.0 & -16.31 & 0.90 & 32 & (16)\\
NGC 1156 & IB &  37.0 & 34.8 & 9.8 &\nodata& 7.0 & -18.00 & 3.75 & 60 & (16)\\
NGC 2366 & IB &  34.5 & 71.2 & 9.8 &\nodata& 3.2 & -16.69 & 2.44 & 47 & (16)\\
NGC 4214 & I  &  33.0 & 22.7 & 9.8 &\nodata& 4.8 & -18.58 & 5.95 & 50 & (16)
\enddata
\tablerefs{(1) \citet{Tsv95}; (2) \citet{Bra06}; (3) \citet{Pel78};
(4) \citet{Hod99}; (5) \citet{Eva96}; (6) \citet{Fei97}; (7) \citet{Lin03};
(8) \citet{Gon97}; (9) \citet{Roz00}; (10) \citet{Cou93}; (11) \citet{Pet96};
(12) \citet{Hod90}; (13) \citet{Kna93}; (14) \citet{Pet98}; (15) \citet{Roz99};
(16) \citet{Roy00}.}
\end{deluxetable}

\clearpage

\begin{deluxetable}{lrcccclrccc}
\tablecolumns{11}
\tablewidth{0pt}
\tablecaption{Calculated fractal dimensions\label{table3}}
\tablehead{
\colhead{Name} & \colhead{$N$} & \colhead{$D_c$} &
\colhead{$\sigma_{boot}$} & \colhead{$R_{min}-R_{max}$} & &
\colhead{Name} & \colhead{$N$} & \colhead{$D_c$} &
\colhead{$\sigma_{boot}$} & \colhead{$R_{min}-R_{max}$}
}
\startdata
\cutinhead{Spirals}
ESO 377-24 & 59 & 1.92 & 0.42 & 0.20 - \phantom{0}2.82 & &
IC 2510  &   70 & 1.86 & 0.22 & 0.23 - \phantom{0}2.51 \\
IC 2560  &  137 & 1.74 & 0.09 & 0.35 - 10.14 & & 
IC 3639  &  112 & 2.11 & 0.09 & 0.11 - \phantom{0}3.99 \\
NGC 210  &  162 & 1.52 & 0.08 & 0.16 - 11.93 & & 
NGC 224  &  981 & 1.51 & 0.02 & 0.01 - 17.38 \\
NGC 337A &  189 & 1.34 & 0.05 & 0.05 - \phantom{0}9.93 & & 
NGC 598  & 1272 & 1.83 & 0.02 & 0.01 - \phantom{0}5.71 \\
NGC 628  & 2027 & 1.84 & 0.01 & 0.03 - \phantom{0}9.27 & & 
NGC 864  &  226 & 1.68 & 0.04 & 0.10 - \phantom{0}6.65 \\
NGC 1042 &  158 & 1.60 & 0.08 & 0.11 - \phantom{0}5.35 & & 
NGC 1068 &  166 & 1.99 & 0.07 & 0.09 - \phantom{0}5.65 \\
NGC 1073 &  170 & 1.41 & 0.07 & 0.02 - \phantom{0}7.93 & & 
NGC 1097 &  402 & 1.78 & 0.03 & 0.01 - 16.30 \\
NGC 1300 &   84 & 1.08 & 0.24 & 0.15 - \phantom{0}8.75 & & 
NGC 1566 &  679 & 1.77 & 0.02 & 0.04 - \phantom{0}9.79 \\
NGC 1672 &  260 & 1.91 & 0.05 & 0.01 - 11.16 & & 
NGC 1808 &  206 & 1.70 & 0.06 & 0.04 - \phantom{0}2.36 \\
NGC 2775 &   66 & 1.27 & 0.28 & 0.16 - \phantom{0}3.79 & & 
NGC 2805 &   94 & 1.60 & 0.11 & 0.15 - 12.05 \\
NGC 2985 &  338 & 1.87 & 0.03 & 0.09 - \phantom{0}8.42 & & 
NGC 2997 &  373 & 1.89 & 0.04 & 0.19 - 11.29 \\
NGC 3031 &  493 & 1.74 & 0.04 & 0.04 - \phantom{0}9.73 & & 
NGC 3081 &   75 & 1.89 & 0.23 & 0.10 - \phantom{0}3.14 \\
NGC 3169 &   67 & 1.35 & 0.26 & 0.14 - \phantom{0}3.63 & & 
NGC 3184 &  576 & 1.70 & 0.02 & 0.04 - \phantom{0}8.51 \\
NGC 3227 &  185 & 1.86 & 0.07 & 0.12 - \phantom{0}5.46 & & 
NGC 3344 &  669 & 1.78 & 0.03 & 0.03 - \phantom{0}6.63 \\
NGC 3351 &   55 & 0.98 & 0.14 & 0.20 - \phantom{0}4.77 & & 
NGC 3359 &  547 & 1.83 & 0.02 & 0.12 - 13.53 \\
NGC 3367 &   79 & 1.74 & 0.19 & 0.13 - \phantom{0}6.91 & & 
NGC 3368 &   77 & 0.86 & 0.21 & 0.06 - \phantom{0}3.74 \\
NGC 3393 &   80 & 1.28 & 0.34 & 0.18 - \phantom{0}7.25 & & 
NGC 3486 &  612 & 1.78 & 0.02 & 0.03 - \phantom{0}7.89 \\
NGC 3631 &  801 & 1.97 & 0.01 & 0.02 - 12.96 & & 
NGC 3660 &   59 & 1.30 & 0.27 & 0.40 - \phantom{0}8.08 \\
NGC 3726 &  614 & 1.80 & 0.03 & 0.05 - \phantom{0}8.94 & & 
NGC 3783 &   58 & 2.07 & 0.27 & 0.21 - \phantom{0}1.99 \\
NGC 3810 &  400 & 2.00 & 0.03 & 0.06 - \phantom{0}5.88 & & 
NGC 3982 &  117 & 2.00 & 0.11 & 0.06 - \phantom{0}1.71 \\
NGC 4030 &  276 & 2.01 & 0.03 & 0.07 - \phantom{0}7.08 & & 
NGC 4051 &  232 & 1.74 & 0.07 & 0.08 - \phantom{0}5.60 \\
NGC 4123 &  247 & 1.66 & 0.04 & 0.06 - \phantom{0}9.90 & & 
NGC 4151 &  262 & 1.53 & 0.03 & 0.09 - 12.97 \\
NGC 4254 &  626 & 1.86 & 0.02 & 0.09 - 18.10 & & 
NGC 4258 &  137 & 1.79 & 0.09 & 0.32 - 12.86 \\
NGC 4303 &  873 & 1.83 & 0.02 & 0.06 - 16.71 & & 
NGC 4321 & 2601 & 1.94 & 0.01 & 0.05 - 19.34 \\
NGC 4395 &  498 & 1.49 & 0.02 & 0.01 - \phantom{0}2.51 & & 
NGC 4487 &  146 & 1.78 & 0.05 & 0.11 - \phantom{0}6.30 \\
NGC 4507 &   92 & 1.62 & 0.21 & 0.21 - \phantom{0}4.24 & & 
NGC 4535 &  518 & 1.71 & 0.02 & 0.09 - 21.16 \\
NGC 4548 &   74 & 1.16 & 0.15 & 0.03 - \phantom{0}1.74 & & 
NGC 4579 &  121 & 1.38 & 0.09 & 0.06 - \phantom{0}6.67 \\
NGC 4593 &  112 & 1.28 & 0.15 & 0.08 - \phantom{0}8.03 & & 
NGC 4602 &  218 & 2.07 & 0.05 & 0.24 - \phantom{0}8.29 \\
NGC 4618 &  290 & 1.76 & 0.05 & 0.06 - \phantom{0}3.70 & & 
NGC 4639 &  190 & 1.89 & 0.06 & 0.01 - \phantom{0}5.18 \\
NGC 4689 &  160 & 2.00 & 0.06 & 0.13 - \phantom{0}5.53 & & 
NGC 4699 &  104 & 1.77 & 0.13 & 0.26 - \phantom{0}4.18 \\
NGC 4725 &  134 & 1.20 & 0.19 & 0.06 - \phantom{0}6.89 & & 
NGC 4736 &  294 & 1.71 & 0.04 & 0.03 - \phantom{0}3.21 \\
NGC 4939 &  250 & 1.86 & 0.04 & 0.24 - 19.39 & & 
NGC 4995 &  142 & 1.91 & 0.09 & 0.04 - \phantom{0}3.99 \\
NGC 5033 &  423 & 2.00 & 0.03 & 0.31 - 15.89 & & 
NGC 5194 &  478 & 2.05 & 0.03 & 0.01 - \phantom{0}7.97 \\
NGC 5247 &  157 & 1.61 & 0.07 & 0.09 - \phantom{0}7.79 & & 
NGC 5248 &  381 & 1.88 & 0.03 & 0.07 - 15.14 \\
NGC 5334 &  106 & 1.57 & 0.18 & 0.10 - \phantom{0}5.76 & & 
NGC 5364 &  174 & 2.04 & 0.05 & 0.32 - 11.62 \\
NGC 5371 &  264 & 1.66 & 0.06 & 0.16 - 14.01 & & 
NGC 5427 &  300 & 1.78 & 0.04 & 0.10 - 10.93 \\
NGC 5457 & 1264 & 1.76 & 0.01 & 0.05 - 25.82 & & 
NGC 5474 &  165 & 1.46 & 0.06 & 0.04 - \phantom{0}3.34 \\
NGC 5643 &  214 & 1.77 & 0.05 & 0.04 - \phantom{0}4.79 & & 
NGC 5681 &   55 & 2.06 & 0.38 & 2.55 - 22.09 \\
NGC 5850 &  155 & 1.31 & 0.10 & 0.20 - 16.99 & & 
NGC 5921 &  274 & 1.77 & 0.04 & 0.08 - 12.77 \\
NGC 5964 &  111 & 1.55 & 0.13 & 0.09 - \phantom{0}8.99 & & 
NGC 6070 &   62 & 2.00 & 0.31 & 0.69 - \phantom{0}6.98 \\
NGC 6118 &  119 & 2.09 & 0.11 & 0.41 - 10.81 & & 
NGC 6140 &  127 & 1.58 & 0.07 & 0.08 - \phantom{0}7.79 \\
NGC 6221 &  173 & 1.70 & 0.06 & 0.08 - \phantom{0}4.83 & & 
NGC 6300 &  317 & 1.97 & 0.05 & 0.01 - \phantom{0}7.45 \\
NGC 6384 &  283 & 2.03 & 0.04 & 0.38 - 16.84 & & 
NGC 6814 &  734 & 1.87 & 0.02 & 0.07 - \phantom{0}8.05 \\
NGC 6946 & 1528 & 1.75 & 0.01 & 0.02 - \phantom{0}8.00 & & 
NGC 7314 &  151 & 2.06 & 0.09 & 0.01 - \phantom{0}7.02 \\
NGC 7331 &  252 & 1.90 & 0.05 & 0.09 - 15.51 & & 
NGC 7479 & 1009 & 1.70 & 0.02 & 0.08 - 18.07 \\
NGC 7552 &   78 & 1.88 & 0.13 & 0.29 - \phantom{0}5.55 & & 
NGC 7590 &  129 & 2.01 & 0.10 & 0.07 - \phantom{0}3.30 \\
NGC 7741 &  246 & 1.66 & 0.05 & 0.06 - \phantom{0}5.54 \\
\cutinhead{Irregulars}
DDO 47   &  57 & 1.63 & 0.21 & 0.05 - 1.42 & & 
DDO 50   &  89 & 1.56 & 0.10 & 0.06 - 1.72 \\
DDO 154  &  55 & 1.83 & 0.14 & 0.06 - 0.70 & & 
DDO 168  &  49 & 2.09 & 0.20 & 0.07 - 0.64 \\
IC 10    &  79 & 1.64 & 0.14 & 0.03 - 0.48 & & 
NGC 1156 &  95 & 2.07 & 0.11 & 0.12 - 1.70 \\
NGC 2366 &  97 & 1.75 & 0.18 & 0.06 - 2.22 & & 
NGC 4214 & 186 & 1.73 & 0.05 & 0.06 - 3.88
\enddata
\end{deluxetable}

\clearpage

\begin{deluxetable}{lccccccc}
\tablecolumns{8}
\tablewidth{0pt}
\tablecaption{Summary of calculated fractal dimensions\label{table4}}
\tablehead{
\colhead{} & \multicolumn{3}{c}{$D_c$} &
\colhead{} & \multicolumn{3}{c}{$\sigma_{boot}$} \\
\cline{2-4} \cline{6-8} \\
\colhead{HII regions} &
\colhead{mean} & \colhead{median} & \colhead{sta.dev.} & &
\colhead{mean} & \colhead{median} & \colhead{sta.dev.}
}
\startdata
$\phantom{000 <}N<100$ & 1.62 & 1.64 & 0.37 & & 0.22 & 0.21 & 0.08\\
$100 < N < 175$        & 1.71 & 1.72 & 0.28 & & 0.09 & 0.09 & 0.04\\
$175 < N < 375$        & 1.79 & 1.77 & 0.16 & & 0.05 & 0.05 & 0.01\\
$375 < N$              & 1.81 & 1.80 & 0.13 & & 0.02 & 0.02 & 0.01
\enddata
\end{deluxetable}

\clearpage

\begin{deluxetable}{lccccccccc}
\tablecolumns{9}
\tablewidth{0pt}
\tablecaption{Fractal dimensions
for different HII region brightness\label{table5}}
\tablehead{
\colhead{} & \colhead{} & \multicolumn{2}{c}{High-brightness} & 
\colhead{} & \multicolumn{2}{c}{Medium-brightness} &
\colhead{} & \multicolumn{2}{c}{Low-brightness} \\
\cline{3-4} \cline{6-7} \cline{9-10} \\
\colhead{Galaxy} & &
\colhead{$D_c$} & \colhead{$\sigma_{boot}$} & &
\colhead{$D_c$} & \colhead{$\sigma_{boot}$} & &
\colhead{$D_c$} & \colhead{$\sigma_{boot}$} 
}
\startdata
NGC~628  & & 1.75 & 0.02 & & 1.87 & 0.02 & & 1.83 & 0.02\\
NGC~3344 & & 1.76 & 0.04 & & 1.87 & 0.05 & & 1.89 & 0.05\\
NGC~3486 & & 1.74 & 0.03 & & 1.77 & 0.06 & & 1.81 & 0.07\\
NGC~3631 & & 1.84 & 0.04 & & 1.90 & 0.05 & & 1.94 & 0.04\\
NGC~3726 & & 1.63 & 0.07 & & 1.75 & 0.06 & & 1.92 & 0.06\\
NGC~4254 & & 1.88 & 0.05 & & 1.79 & 0.06 & & 1.70 & 0.06\\
NGC~4303 & & 1.83 & 0.04 & & 1.82 & 0.04 & & 1.83 & 0.04\\
NGC~4321 & & 1.79 & 0.02 & & 1.89 & 0.02 & & 1.83 & 0.02\\
NGC~6946 & & 1.64 & 0.03 & & 1.82 & 0.03 & & 1.79 & 0.03
\enddata
\end{deluxetable}


\clearpage

\clearpage

\clearpage

\clearpage
{\plotone{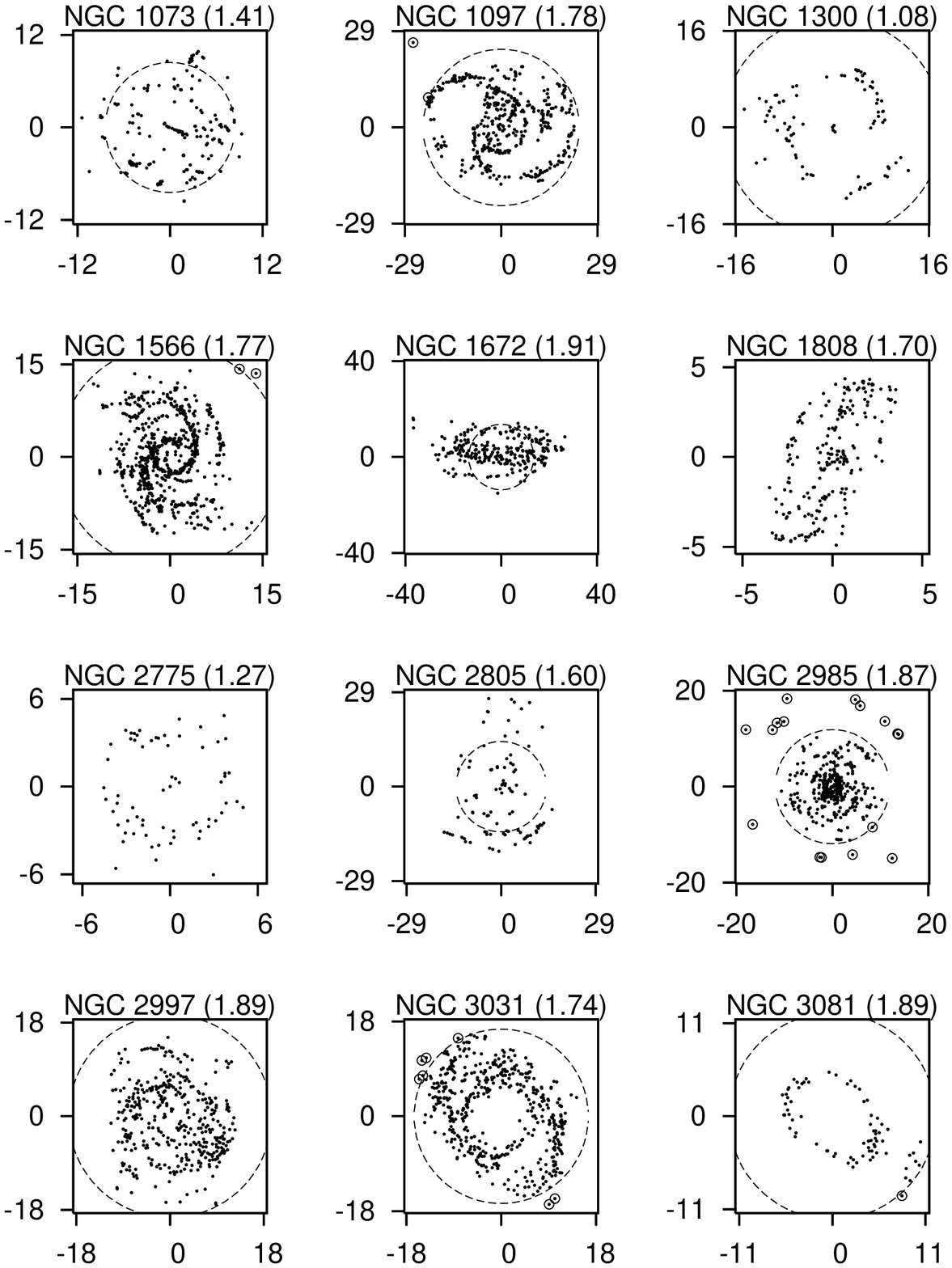}}\\
\centerline{Fig. 3. --- Continued.}
\clearpage
{\plotone{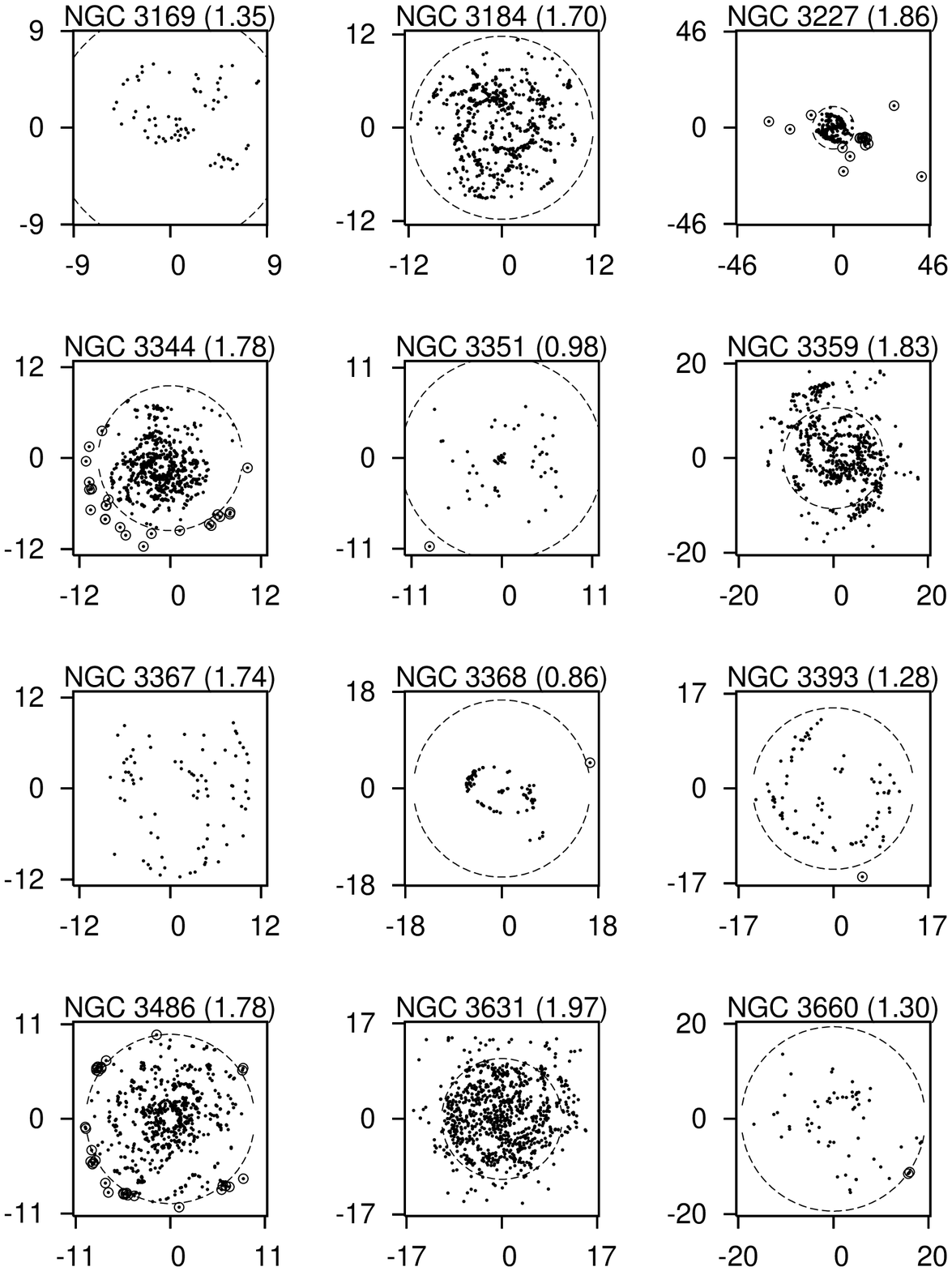}}\\
\centerline{Fig. 3. --- Continued.}
\clearpage
{\plotone{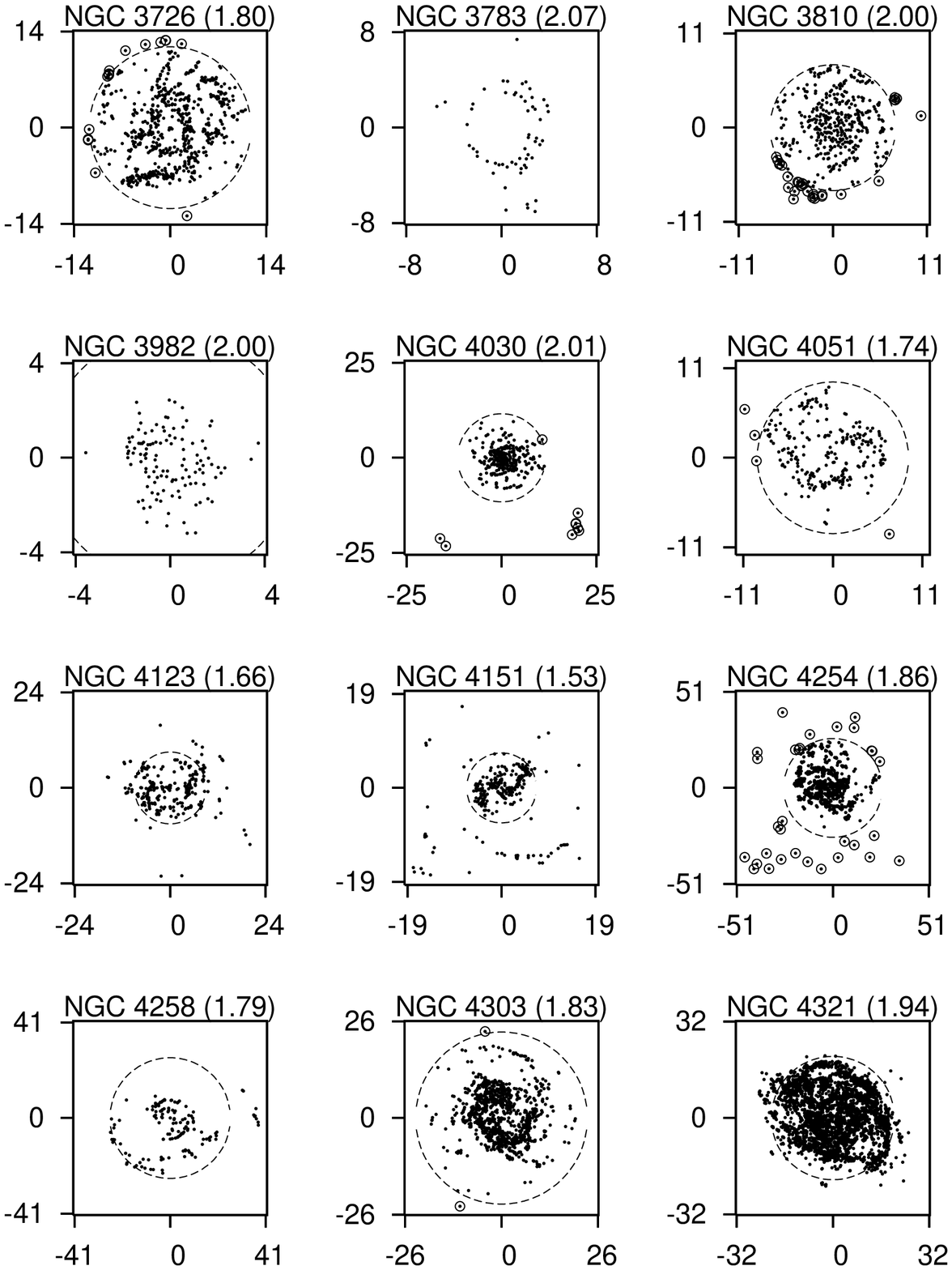}}\\
\centerline{Fig. 3. --- Continued.}
\clearpage
{\plotone{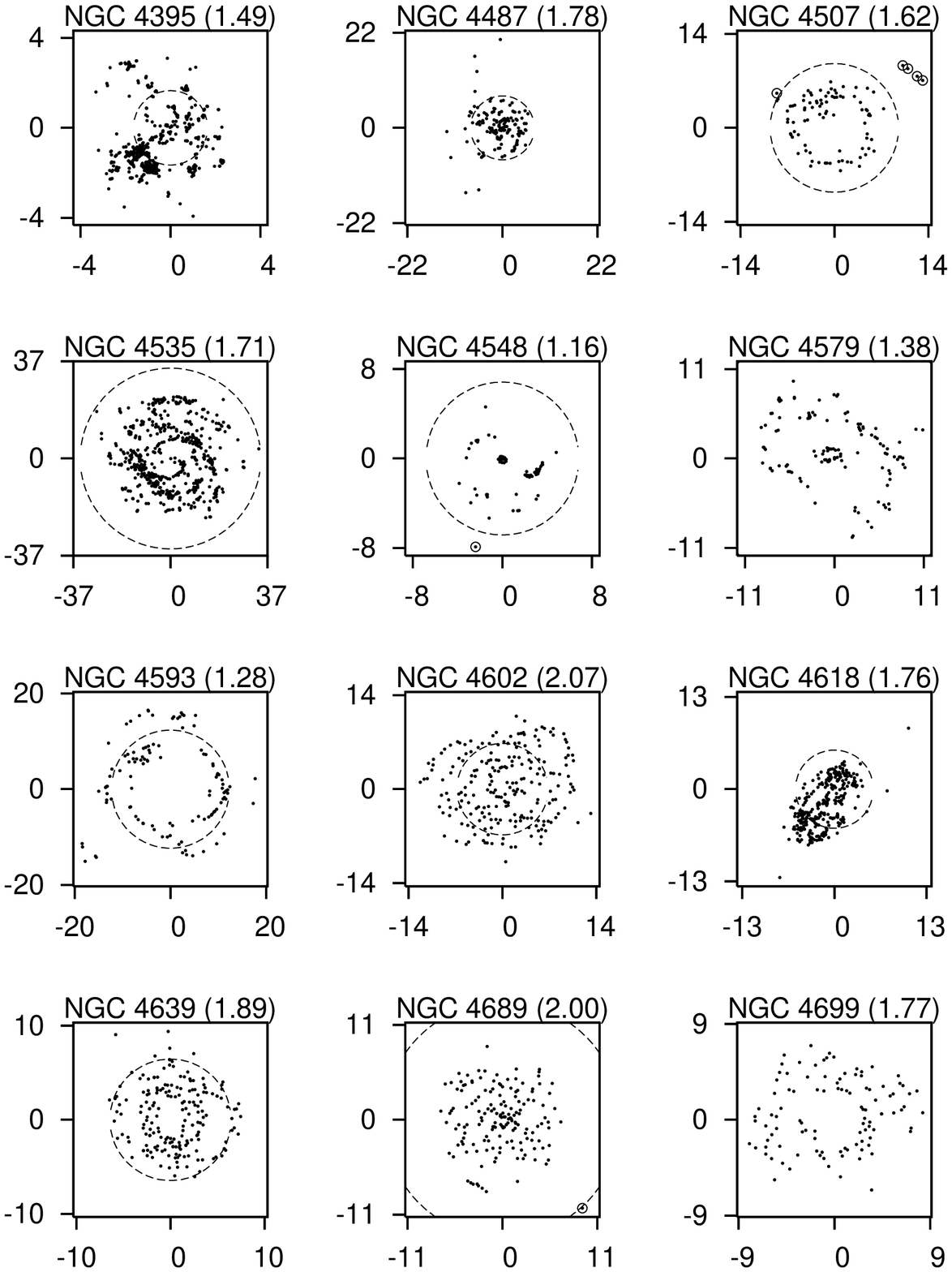}}\\
\centerline{Fig. 3. --- Continued.}
\clearpage
{\plotone{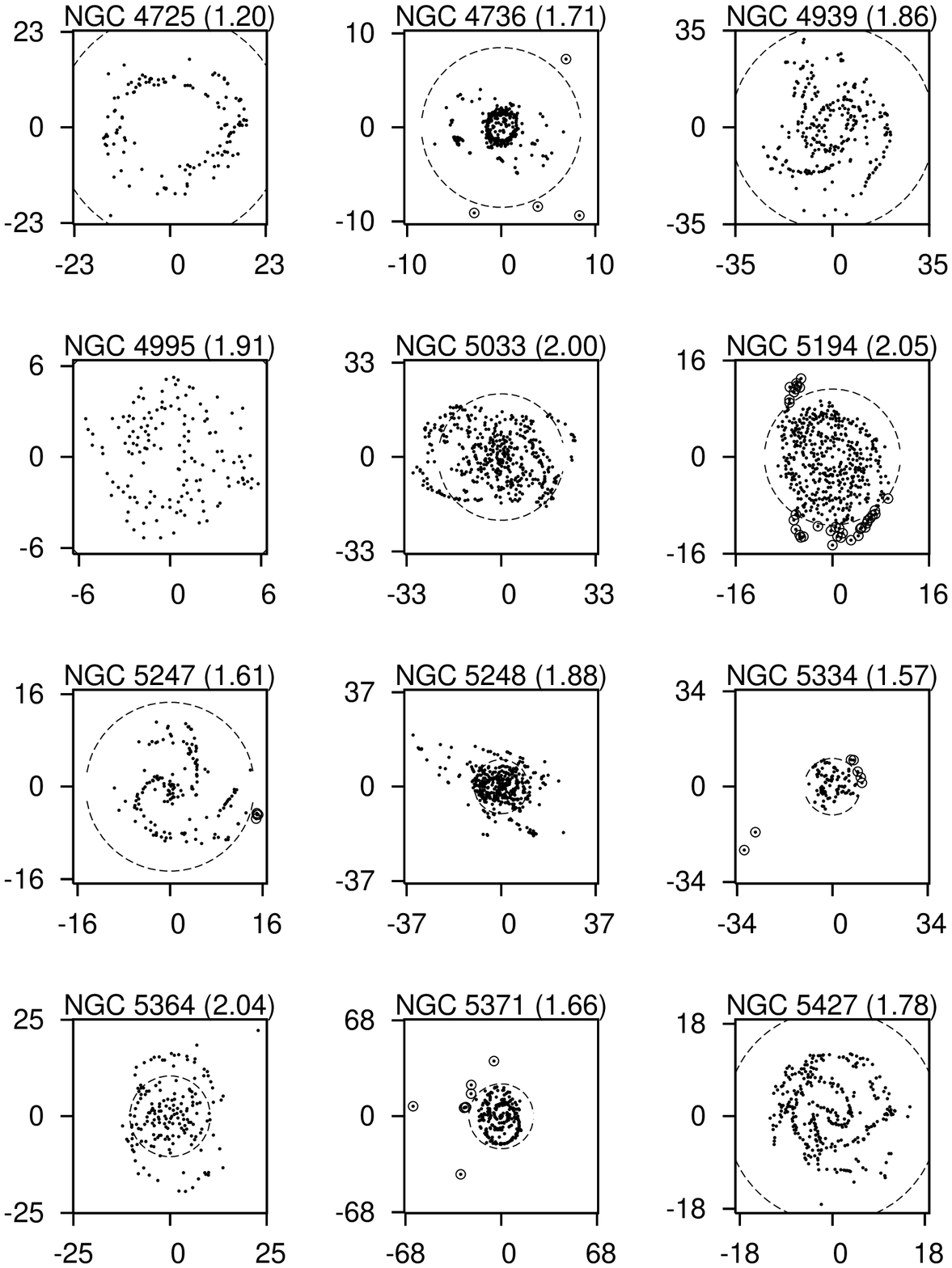}}\\
\centerline{Fig. 3. --- Continued.}
\clearpage
{\plotone{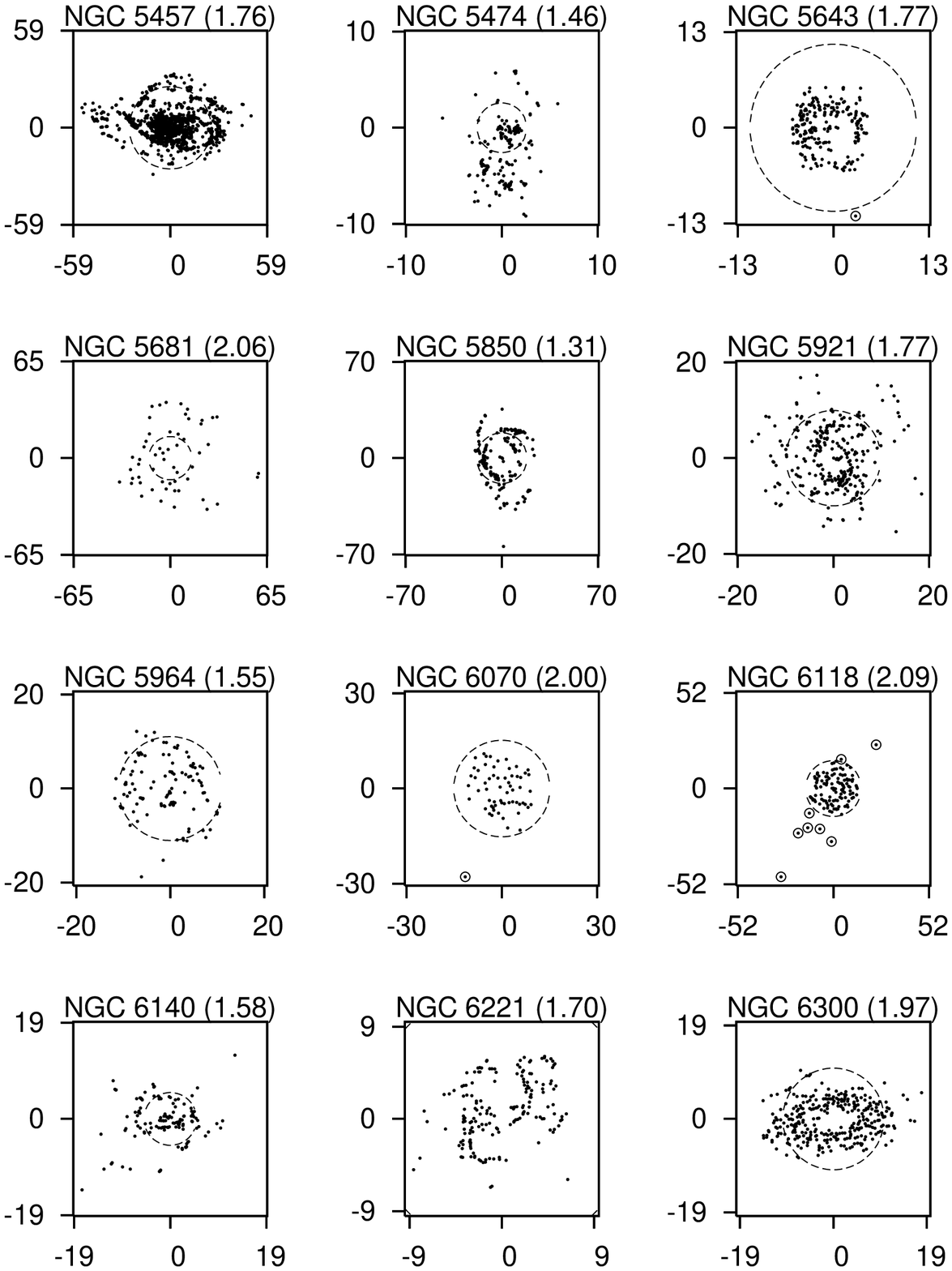}}\\
\centerline{Fig. 3. --- Continued.}
\clearpage
{\plotone{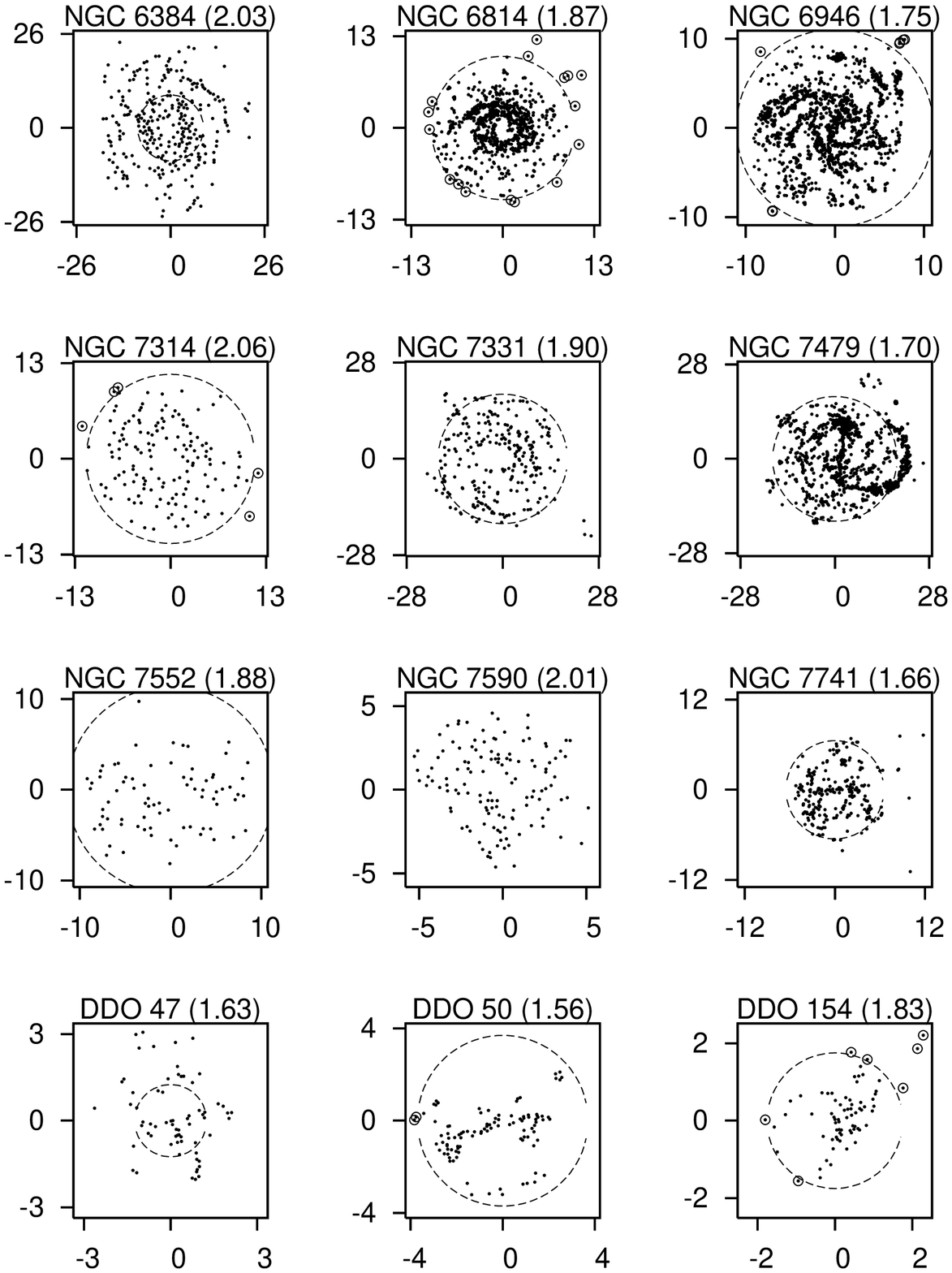}}\\
\centerline{Fig. 3. --- Continued.}
\clearpage
{\plotone{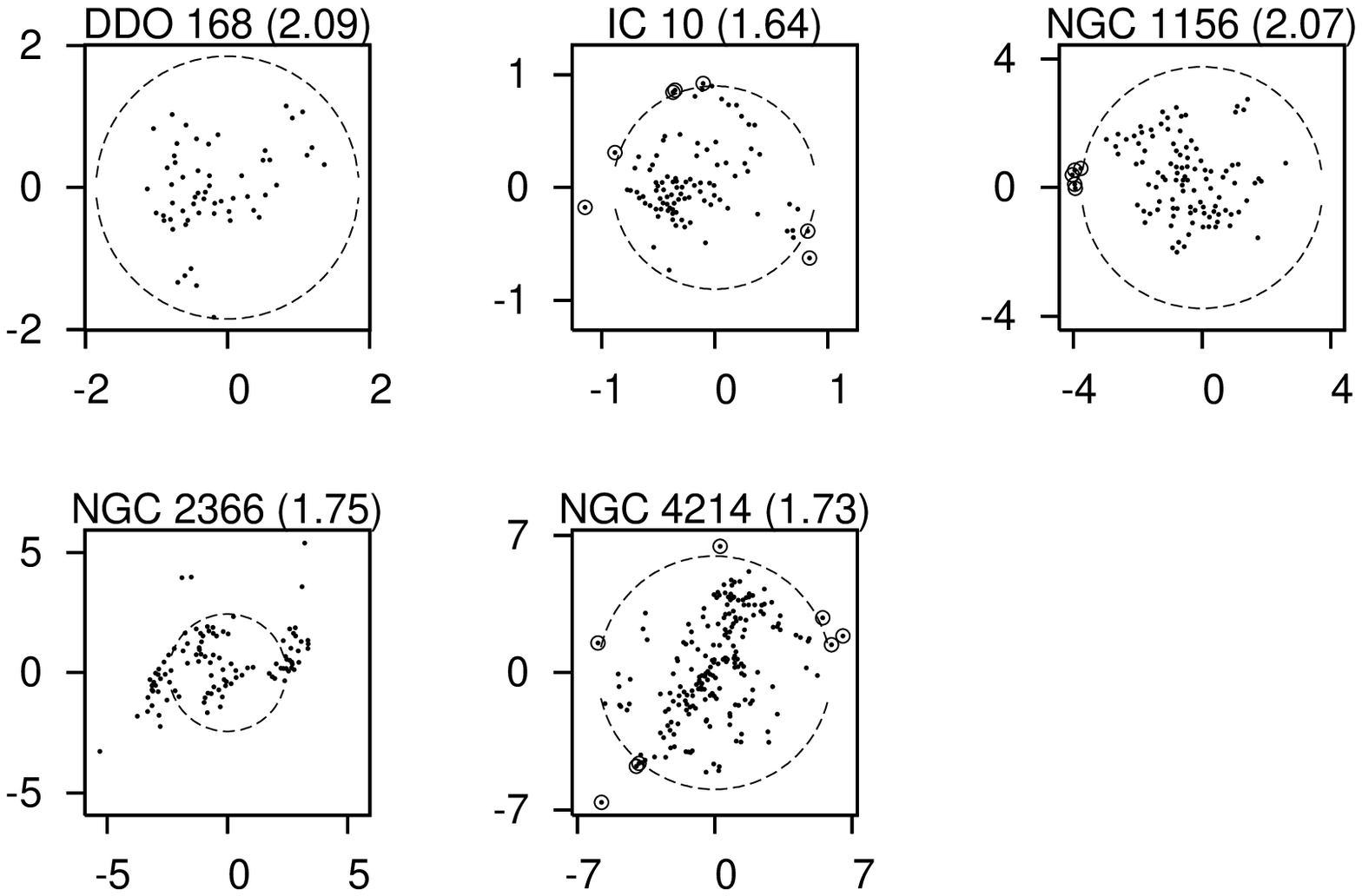}}\\
\centerline{Fig. 3. --- Continued.}

\clearpage

\clearpage

\begin{thebibliography}{}
\bibitem[Akaike(1974)]{Aka74} Akaike, H. 1974,
         IEEE Transactions on Automatic Control, 19, 716
\bibitem[Bastian et al.(2007)]{Bas07} Bastian, N., Ercolano, B., Gieles,
         M., Rosolowsky, E., Scheepmaker, R.~A., Gutermuth, R., \&
         Efremov, Y. 2007, \mnras, 379, 1302 
\bibitem[Begun et al.(2006)]{Beg06}
         Begun, A., Chengalur, J. N., \& Bhardwaj, S. 2006,
         \mnras, 372, L33
\bibitem[Bergin \& Tafalla(2007)]{Ber07}
         Bergin, E. A. \& Tafalla, M. 2007, \araa 45, 339
\bibitem[Bradley et al.(2006)]{Bra06}
         Bradley, T. R., Knapen, J. H., Beckman, J. E., \& Folkes S. L.
         2006, \aap, 459, L13
\bibitem[Chappell \& Scalo(2001)]{Cha01}
         Chappell, D. \& Scalo, J. 2001, \apj, 551, 712
\bibitem[Courtes et al.(1993)]{Cou93}
         Courtes, G., Petit, H., Hua, C. T., Martin, P., Blecha, A.,
         Huguenin, D., \& Golay, M. 1993, \aap, 268, 419
\bibitem[de la Fuente Marcos \& de la Fuente Marcos(2006)]{Fue06}
         de la Fuente Marcos, R., \& de la Fuente Marcos, C. 2006,
         \mnras, 372, 279 
\bibitem[Dutta et al.(2008)]{Dut08} Dutta, P., Begum, A., Bharadwaj, S.,
         \& Chengalur, J. N. 2008, \mnras, 384, L34
\bibitem[Eddy(1977)]{Edd77}
         Eddy, W. F. 1977, {\it Trans. Math. Soft.}, 3, 398
\bibitem[Elmegreen \& Elmegreen(1987)]{Elm87}
         Elmegreen, D. M. \& Elmegreen, B. G. 1987, \apj, 314, 3
\bibitem[Elmegreen \& Elmegreen(1986)]{Elm86} Elmegreen, B. G., \&
         Elmegreen, D. M. 1986, \apj, 311, 554
\bibitem[Elmegreen \& Elmegreen(2001)]{Elm01}
         Elmegreen, B. G. \& Elmegreen, D. M. 2001, \aj, 121, 1507
\bibitem[Elmegreen et al.(2006)]{Elm06} Elmegreen, B. G., Elmegreen, D.
         M., Chandar, R., Whitmore, B., \& Regan, M. 2006, \apj, 644, 879 
\bibitem[Elmegreen et al.(2003)]{Elm03} Elmegreen, B. G., Elmegreen, D.
         M., \& Leitner, S. N. 2003, \apj, 590, 271 
\bibitem[Elmegreen \& Scalo(2004)]{Elm04}
         Elmegreen, B. G. \& Scalo, J. 2004, \araa 42, 211
\bibitem[Evans et al.(1996)]{Eva96}
         Evans, I. N., Koratkar, A. P., Storchi-Bergmann, T., Kirkpatrick,
         H., Heckman, T. M., \& Wilson A. S. 1996, \apjs, 105, 93
\bibitem[Falconer(1990)]{Fal90}
         Falconer, K. J. 1990, Fractal Geometry: Mathematical
         Foundations and Applications (London: Wiley)
\bibitem[Federrath et al.(2007)]{Fed07}
         Federrath, C., Schmidt, W., \& Klessen, R. S. 2007,
         preprint (arXiv:0710.1359)
\bibitem[Feinstein(1997)]{Fei97}
         Feinstein, C. 1997, \apjs, 112, 29
\bibitem[Feitzinger \& Galinski(1987)]{Fei87}
         Feitzinger, J. V., \& Galinski, T. 1987, \aap, 179, 249
\bibitem[Garcia-Gomez \& Athanassoula(1991)]{Gar91}
         Garcia-Gomez, C., \& Athanassoula, E. 1991,
         \aaps, 89, 159
\bibitem[Garcia-Gomez et al.(2002)]{Gar02}
         Garcia-Gomez, C., Athanassoula, E., \& Barbera, C. 2002,
         \aap, 389, 68
\bibitem[Gonzalez-Delgado et al.(1997)]{Gon97}
         Gonzalez-Delgado, R. M., Perez, E., Tadhunter, C., Vilchez,
         J. M., \& Rodriguez-Espinosa, J. M. 1997, \apjs, 108, 155
\bibitem[Grassberger \& Procaccia(1983)]{Gra83}
         Grassberger, P., \& Procaccia, I. 1983, \prl, 50, 346
\bibitem[Hodge(1985)]{Hod85} Hodge, P. 1985, \pasp, 97, 688 
\bibitem[Hodge et al.(1990)]{Hod90}
         Hodge, P., Gurwell, M., Goldader J. D., \& Kennicutt, R. C.
         Jr. 1990, \apjs, 73, 661
\bibitem[Hodge et al.(1999)]{Hod99}
         Hodge, P. W., Balsley, J., Wyder, T. K., \& Skelton, B. P.
         1999, \pasp, 111, 685
\bibitem[Kim et al.(2003)]{Kim03}
         Kim, S., Staveley-Smith, L., Dopita, M. A., Sault, R. J.,
         Freeman, K. C., Lee, Y., \& Chu, Y.-H. 2003, \apjs, 148, 473
\bibitem[Knapen et al.(1993)]{Kna93}
         Knapen, J. H., Arnth-Jensen, N., Cepa, J., \& Beckman, J. E.
         1993, \aj, 106, 56
\bibitem[Lin et al.(2003)]{Lin03}
         Lin, W., Zhou, X., Burstein, D., Windhorst, R. A., Chen, J.,
         Chen, W.-P., Jiang, Z., Kong, X., Ma, J., Sun, W.-H., Wu H.,
         Xue, S., \& Zhu J. 2003, \aj, 126, 1286
\bibitem[Ochsenbein et al.(2000)]{Och00}
         Ochsenbein, F., Bauer, P., Marcout, J. 2000, \aaps, 143, 221
\bibitem[Odekon(2006)]{Ode06}
         Odekon, M. C. 2006, \aj, 132, 1834
\bibitem[Oey \& Clarke(1998)]{Oey98}
         Oey, M. S. \& Clarke, C. J. 1998, \aj, 115, 1543
\bibitem[Paladini et al.(2004)]{Pal04}
         Paladini, R., Davies, R. D., \& DeZotti, G. 2004,
         \mnras, 347, 237
\bibitem[Parodi \& Binggeli(2003)]{Par03}
         Parodi, B. R., \& Binggeli, B. 2003, \aap, 398, 501
\bibitem[Paturel et al.(2003)]{Pat03}
         Paturel, G., Petit, C., Prugniel, Ph., Theureau, G., Rousseau,
         J., Brouty, M., Dubois, P., \& Cambresy, L. 2003, \aap, 412, 45
\bibitem[Pellet et al.(1978)]{Pel78}
         Pellet, A., Astier, N., Viale, A., Courtes, G., Maucherat, A.,
         Monnet, G., \& Simien, F. 1978, \aaps, 31, 439
\bibitem[Petit et al.(1996)]{Pet96}
         Petit, H., Hua, C. T., Bersier, D., \& Courtes G. 1996,
         \aap, 309, 446
\bibitem[Petit(1998)]{Pet98}
         Petit, H. 1998, \aaps, 131, 317
\bibitem[Pfenniger \& Combes(1994)]{Pfe94} Pfenniger, D., \& Combes,
         F. 1994, \aap, 285, 94 
\bibitem[R Development Core Team(2008)]{R08}
         R Development Core Team 2008,
         {\it R: A Language and Environment for Statistical Computing},
         R Foundation for Statistical Computing, Vienna, Austria,
         URL http://www.R-project.org, ISBN 3-900051-07-0
\bibitem[Roye \& Hunter(2000)]{Roy00}
         Roye, E. W., \& Hunter, D. A. 2000, \aj, 119, 1145
\bibitem[Rozas et al.(1999)]{Roz99}
         Rozas, M., Zurita, A., Heller, C. H., \& Beckman, J. E. 1999,
         \aaps, 135, 145
\bibitem[Rozas et al.(2000)]{Roz00}
         Rozas, M., Zurita, A., \& Beckman, J. E. 2000, \aap, 354, 823
\bibitem[Silk(1997)]{Sil97} Silk, J. 1997, \apj, 481, 703 
\bibitem[S\'anchez et al.(2005)]{San05}
         S\'anchez, N., Alfaro, E. J., \& P\'erez, E. 2005,
         \apj, 625, 849
\bibitem[S\'anchez et al.(2007a)]{San07GB}
         S\'anchez, N., Alfaro, E. J., Elias, F., Delgado, A. J., \&
         Cabrera-Ca\~no, J. 2007a, \apj, 667, 213
\bibitem[S\'anchez et al.(2007b)]{San07Df}
         S\'anchez, N., Alfaro, E. J., \& P\'erez, E. 2007b,
         \apj, 656, 222
\bibitem[Stanimirovic et al.(1999)]{Sta99}
         Stanimirovic, S., Staveley-Smith, L., Dickey, J. M., Sault,
         R. J., \& Snowden, S. L. 1999, \mnras, 302, 417
\bibitem[Tassis(2007)]{Tas07} Tassis, K. 2007, \mnras, 382, 1317 
\bibitem[Tsvetanov \& Petrosian(1995)]{Tsv95}
         Tsvetanov, Z. I., \& Petrosian A. R. 1995, \apjs, 101, 287
\bibitem[Westpfahl et al.(1999)]{Wes99}
         Westpfahl, D. J., Coleman, P. H., Alexander, J., \& Tongue,
         T. 1999, \aj, 117, 868
\bibitem[Willett et al.(2005)]{Wil05} Willett, K. W., Elmegreen, B. G.,
         \& Hunter, D. A.\ 2005, \aj, 129, 2186
\end{thebibliography}
\end{document}